\documentclass[aps,pre,twocolumn,groupedaddress]{revtex4-2}
\usepackage{amsmath}
\usepackage{amssymb}
\usepackage{graphics}
\usepackage{graphicx}
\usepackage{hyperref}
\usepackage{color}
\usepackage{bm}
			
\usepackage{soul}
\usepackage[dvipsnames]{xcolor}

\newcommand{\1}{1}
\def\<{\langle}
\def\>{\rangle}
\def\Ttr{{T_{\rm tr}}}
\def\Tte{{T_{\rm te}}}
\def\Tva{{T_{\rm va}}}
\def\Tin{{T_{\rm in}}}
\def\Xtr{{X^{\rm (tr)}}}
\def\Xte{{X^{\rm (te)}}}
\def\Xin{{X^{\rm (in)}}}
\def\Xva{{X^{\rm (va)}}}
\def\Ns{{N_\text{s}}}
\def\E{E}
\def\N{{\cal N}}
\def\Pv{P^{{\rm v}}}
\def\Cv{{C^{{\rm v}}}}
\def\Jv{J^{{\rm v}}}

\def\x{{\bf x}}
\def\y{{\bf y}}

\def\traza{{\rm tr}}

\def\myi{\imath}

\def\L{{\cal L}}

\begin{document}


\title{Noise-cleaning the precision matrix of fMRI time series}


\author{Miguel Ibáñez-Berganza}
\affiliation{Istituto Italiano di Tecnologia. Largo Barsanti e Matteucci, 53, 80125 Napoli, Italy}
\author{Carlo Lucibello}
\affiliation{AI Lab, Institute for Data Science and Analytics, Bocconi
	University, 20136 Milano, Italy}
\author{Francesca Santucci}
\affiliation{Networks Unit, IMT School for Advanced Studies Lucca, Piazza San Francesco 19, 50100, Lucca, Italy}
\author{Tommaso Gili}
\affiliation{Networks Unit, IMT School for Advanced Studies Lucca, Piazza San Francesco 19, 50100, Lucca, Italy}
\author{Andrea Gabrielli}
\affiliation{Dipartimento di Ingegneria Civile, Informatica e delle Tecnologie Aeronautiche, Università degli Studi “Roma Tre”, Via Vito Volterra 62, 00146 Rome, Italy}
\affiliation{Centro Ricerche “Enrico Fermi”, Via Panisperna 89a, 00184 Rome, Italy}


\date{\today}

\begin{abstract}
We present a comparison between various algorithms of inference of covariance and precision matrices in small datasets of real vectors, of the typical length and dimension of human brain activity time series retrieved by functional Magnetic Resonance Imaging (fMRI). Assuming a Gaussian model underlying the neural activity, the problem consists in denoising the empirically observed matrices in order to obtain a better estimator of the (unknown) true precision and covariance matrices.
	We consider several standard noise-cleaning algorithms and compare them on two types of datasets. The first type are time series of fMRI brain activity of human subjects at rest. The second type are synthetic time series sampled from a generative Gaussian model of which we can vary the fraction of dimensions per sample $q=N/T$ and the strength of off-diagonal correlations. The reliability of each algorithm is assessed in terms of test-set likelihood and, in the case of synthetic data, of the distance from the true precision matrix. 
	 We observe that the so called Optimal Rotationally Invariant Estimator, based on Random Matrix Theory, leads to a significantly lower distance from the true precision matrix in synthetic data, and higher  test  likelihood in natural fMRI data. 
	 We propose a variant of the Optimal Rotationally Invariant Estimator in which one of its parameters is optimised by cross-validation. In the severe undersampling  regime (large $q$) typical of fMRI series, it  outperforms all the other estimators. We furthermore propose a simple algorithm based on an iterative likelihood gradient ascent, providing an accurate estimation for weakly correlated datasets.
\end{abstract}


\maketitle


\section{Introduction\label{sec:intro}}

Multiple, complex, rapidly changing brain activity patterns are constrained by an underlying Structural Connectivity (SC) network of neuronal fiber bundles that evolve on distinctly larger time scales \cite{cabral2017,honey2009,vandenheuvel2009,friston2011}. It is precisely this degeneration (reminiscent of collective phases in physics) of emerging complex and segregated dynamical states, subtended by a relatively static and sparse SC network, that makes possible context-sensitive conscious cognition, perception, and action \cite{park2013,demertzi2019,grecius2003}. For this reason, the study of the relation between brain structure and the associated emergent cognitive functions in healthy and diseased subjects is, arguably, one of the most important challenges in neuroscience. Since the advent of high-quality  functional Magnetic Resonance Imaging (fMRI) and Electro/Magneto-Encephalography (EEG/MEG) datasets, a wide range of linear \cite{honey2009,gu2017,gilson2016,liegeois2020} and non-linear \cite{morone2017,fortel2019,watanabe2013,fortel2022,niu2019,abeyasinghe2015,kadirvelu2017,hahn2019,das2014,deco2012}, \cite{wang2019} models linking emergent function and the subtending SC have been considered.

In this context, the role played by {\it correlation} and {\it precision matrices} of brain activity patterns is receiving increasing interest \cite{deligianni2011,liegeois2020,pervaiz2020,liu2020brain,smith2013functional,dadi2019,chung2021,rahim2019,ryali2012,smith2013functional,brier2015}. Functional Connectivity (FC) is usually estimated through the correlation matrix ($C$) among pairs of functional time series of activity (usually BOLD fMRI signals {\it at rest}) corresponding to different brain areas. A related quantity, considered as an alternative estimator of FC (see, for example, \cite{liegeois2020}), is the {\it precision matrix}, or the inverse of the correlation matrix $J=C^{-1}$. Assuming that the vector of brain activity patterns obeys Gaussian statistics \cite{honey2009,gu2017,gilson2016,liegeois2020} (or that the time series follow an Ornstein-Uhlenbeck process \cite{vandenheuvel2009,lefortbesnard}), the precision matrix $J$ represents harmonic coupling constants that constraint the emergent correlations $C$. For this reason, $J$ may be understood as an (inferred) model of more ``direct'' \cite{liegeois2020} anatomical connections between brain areas. Indeed, differently from $C$, and within a Gaussian approximation, $J$ accounts for direct causal relations only. As a matter of fact, in the context of the SC/FC relationship, the inferred precision matrix $J$ has been demonstrated to be a more accurate statistical estimator of SC (as retrieved e.g. by Diffusion Tensor Imaging techniques) than $C$ and than the inter-area couplings resulting from Granger Causality and auto-regressive inference (\cite{deligianni2011,liegeois2020} and references therein). Furthermore, $J$ has also been shown to provide better prediction scores than correlation-based FC for some diseases and non-imaging phenotypic measures \cite{pervaiz2020,liu2020brain,smith2013functional,dadi2019} (see also \cite{chung2021}), and to better capture  intra-subject FC differences \cite{rahim2019}. Finally, recent results suggest \cite{liegeois2020} that the relation between the empirical SC matrix and the precision matrix from temporal BOLD series can be exploited, beyond the Ornstein-Uhlenbeck hypothesis, to infer the relative timescales of temporal correlation of the BOLD activity in different brain sub-networks of grey matter, that in turn reflect the relative complexity of cognitive functions involved in the cortical hierarchy.

The studies of  precision matrices are, however, severely limited by the limited accuracy of their statistical estimation, due to the short length of  time series \cite{smith2013functional,liegeois2020,brier2015,ryali2012,chung2021}. To overcome this issue, several techniques of statistical inference of the correlation (and, hence, of the precision) matrix have been proposed in the context of network neuroscience. These are: Ledoit-Wolf and Tikhonov regularised precision matrix \cite{mejia2016,brier2015,deligianni2014,pervaiz2020,liegeois2020}; $L_1$-regularisation with or without population priors \cite{varoquaux2010brain,ryali2012}; addition of regularised aggregation to construct the group precision matrix \cite{chung2021}; (non-isotropic) population-shrinkage covariance estimators \cite{rahim2019} (see further references in \cite{chung2021,rahim2019,pervaiz2020}). Accurate strategies for correlation and precision matrix regularisation, or noise-cleaning, are also crucial in order to improve the efficiency of auto-regressive models of causal inference \cite{liegeois2020},  since they rely on the inversion of sample covariance estimators constructed from few data vectors. 

Previous discussion highlights the need for accurate benchmarking of estimators of functional connectivity. In the present contribution we compare standard noise-cleaning methods (such as linear shrinkage and Principal Component Analysis) on correlation matrices derived from short time series, of the typical size and length of fMRI data. Importantly, we add to the comparison recent methods grounded on Random Matrix Theory \cite{bun2017}. Despite such methods have been known in statistical physics and in theoretical finance since a few years, they have not been, to the best of our knowledge, benchmarked nor just employed in the study of the human brain connectome. 
The algorithms are compared in two types of datasets:  {\it synthetic} ones, sampled from a known generative Gaussian model, and two {\it natural} datasets of human resting-state brain activity by fMRI. We evaluate the efficiency of each algorithm on each dataset in terms of the out-of-sample (test) likelihood, that takes into account both variance and bias errors, and of two related criteria.
In the case of  synthetic  datasets, the methods' efficiency is further evaluated in terms of the element-wise distance from the {\it true}, or {\it population} precision matrix. Correlation and precision matrices are here independently inferred for each multivariate time-series (each subject). We do not leverage the group-wide information across subjects in our algorithms, which is a relevant direction for further investigations and comparisons.

The article is structured as follows. In Section \ref{sec:methods} we define the problem of noise-cleaning correlation matrices and set the notation. Then we describe the benchmarked algorithms (Section \ref{sec:algorithms}), the quality criteria (Section \ref{sec:metrics}), the cross-validation strategies (Section \ref{sec:crossval}), and the characteristics of the synthetic (Section \ref{sec:generative}) and natural fMRI (Section \ref{sec:fMRIdata}) datasets. We finally present the results and draw the conclusions in Sections \ref{sec:results} and \ref{sec:conclusions} respectively.

\section{Materials and methods \label{sec:methods}}

{\bf Datasets.} Let the dataset $X$ be an $N\times T$ real matrix consisting of $T$ $N$-dimensional vectors, and let $\x(t):=(X_{1t},\ldots,X_{Nt})\in{\mathbb R}^{N}$.
In the context of network neuroscience, $X$ represents the single subject data, $N$ is the number of anatomic areas or regions of interest and $T$ is the length of the time signal. 
We will assume that the observations $\x(t)$ are identically and independently distributed. Furthermore, we assume the signal distribution to be a multivariate Gaussian.
Without loss of generality, we will assume that the data is normalized in such a way that it exhibits null temporal averages and unit standard deviation: $\sum_{t=1}^T x_i(t)=0$, $\sum_{t=1}^T x^2_i(t)=T$. In this article we will focus on the case $T\ge N$, or $q\le 1$ being $q:=N/T$. 

{\bf Sample and population covariance matrices.} We will call $\E=X X^\dag/T$ the {\it sample} correlation matrix. Whenever $T$ is finite or $q=N/T$ is non-negligible, $E$ and its inverse in particular are not good estimators of the (unknown) {\it population} or true ({\it verus}) correlation and precision matrices, $\Cv$ and $\Jv=\Cv^{-1}$, that would have been obtained in the limit of infinitely many data, $T\to\infty$ {\it and} $q\to 0$. From the Marchenko-Pastur Equation (see, for example, \cite{livan2018,bun2017}) one knows, in particular, that the differences in the spectral densities of $E$ and $\Cv$ are governed by $q$ and vanish only for $q=0$. 

{\bf Noise-cleaned estimators of the covariance matrix.} The problem of noise-cleaning a covariance matrix amounts to proposing a noise-cleaned matrix $C$, given $X$, that aims to be as similar as possible to $\Cv$ (and more than what $E$ is) according to some criterion. Equivalently, the cleaned matrix $C$ aims to correct the over-fitting (for small $T$) and the curse of dimensionality (for large $q$) that affect the unbiased estimator $E$. In other words, $C$ should present lower bias$+$variance error \cite{mehta2019}, at the expense of a higher bias error. 

In Bayesian terms, $E$ is the maximum likelihood (ML) estimator of the covariance matrix given the dataset $X$ (assuming a Gaussian likelihood, $\N(X|E)$), while $C$ is a {\it beyond-ML estimator} \cite{mackay2003}, in the sense that it aims at achieving an higher test likelihood, at the expense of a lower train likelihood. 
Indeed, the design of noise-cleaning algorithms can be cast in terms of Bayesian Random Matrix Theory \cite{bun2017}. Most of the algorithms that we consider here depend on a hyper-parameter, generically $\gamma$, such that, for $\gamma=0$ the resulting cleaned matrix is $C_{\gamma=0}=E$ (minimum bias, maximum variance), while for its maximum value $\gamma_+$ it is $C_{\gamma_+}=\1_N$ (maximum bias, minimum variance). The optimal value $\gamma^*$ of the hyper-parameter can be set by cross-validated maximisation of a validation-set likelihood (or by maximisation of the training-set Bayesian evidence, as in PCA-Minka, see before). For infinitely many dataset vectors $T\gg N$, there is no over-fitting and no curse of dimensionality, hence $\gamma^*=0$ and $C=E$.

{\bf Training- and test-sets, and hyper-parameters.} Each subject dataset $X$ is decomposed by columns into training and test datasets, $X=(\Xtr,\Xte)$, of dimensions $N\times \Ttr$, and $N\times \Tte$ respectively. Given $\Xtr$, we will obtain noise-cleaned covariance matrices $C$ from several algorithms, also called {\it methods} in this article. 

Some of the considered methods lead to a cleaned matrix $C_\pi$ depending on an hyper-parameter $\pi$. For these methods, the trainingset is in its turn decomposed into {\it inversion}- (or pure training) and validation-sets, $\Xtr=(\Xin,\Xva)$, of dimensions $\Tin,\Tva$ respectively. The optimal value of the hyper-parameter is chosen by maximisation of a {\it criterion} $Q$, $\pi^*=\arg\max_\pi Q(\Xva|C_\pi)$ evaluated on the validation-set (while $C_\pi$ is computed from the inversion-set), and the inversion-validation split is given by $K$-fold cross-validation. Finally, the {\it quality} of each method according to the criterion $Q$ (cross-validated across random $(\Xin,\Xva)$ partitions when needed) is given by $Q(\Xte|C_{\pi^*})$. We evaluate the average and errors of this quantity, across all the {\it subjects} $X$ belonging to a given {\it collection} of datasets, $(X^{(s)})_s$. 

We describe the considered methods, criteria and datasets in subsections \ref{sec:algorithms}, \ref{sec:metrics}, \ref{sec:generative} and \ref{sec:fMRIdata}, respectively.

\subsection{Algorithms of noise-cleaning correlation matrices \label{sec:algorithms}}

It is out of the scope of this paper to present a complete review of the immense amount of results on the general problem of over-fitting and curse of dimensionality mitigation of covariance matrices. We limit ourselves to mention and compare a list of the better known and most popular algorithms for cleaning correlation matrices according to reference \cite{bun2016}. 
Let the spectral decomposition of the sample correlation matrix $E=\Xtr\Xtr^\dag/\Ttr$ be $E=U^\dag\Lambda U$, where $U$ is orthogonal and $\Lambda$ is a diagonal, real matrix. The noise-cleaned or regularised matrix will be called $C$ and its spectral decomposition $C=W^\dag \hat\Lambda W$ where, again, $W$ is orthogonal and $\hat\Lambda$ diagonal. We will assume their eigenvalues $\lambda_i=\Lambda_{ii} \geq 0$ and $\hat\lambda_i=\hat\Lambda_{ii}$ to be in decreasing order. 
As we will see, most of the standard algorithms modify only the sample spectra, so that $W=U$. 

\paragraph{\bf Eigenvalue Clipping,} or {\bf Principal Component Analysis} ({\bf PCA}), according to which only $p$ eigenvalues of $E$ are considered to be significant, with $0\le p \leq N$. 
\begin{itemize}
\item The cleaned spectrum is set equal to the sample spectrum $\hat\lambda_i = \lambda_i$ whenever $i \le p$, otherwise it is set to a common ``noise'' value: $\hat\lambda_{i>p} = \bar \lambda_p = \sum_{j>p}\lambda_j/(N-p)$, equal to the average of the $N-p$ neglected eigenvalues. 
\item The resulting cleaned matrix is $C=U^\dag \hat\Lambda\ U$. Given $p,\Lambda,U$, the choice of $\bar \lambda$ corresponds to a maximum likelihood prescription (see for example \cite{minka2000}).
\end{itemize}

In PCA, the $\gamma$ hyper-parameter is the number of non-fitted principal components $\gamma=N-p$.

We have implemented two variants of this method: For the first one ({\bf PCA (CV)}), the value of $p$ is set by cross-validation (see \ref{sec:crossval}); In the second one ({\bf PCA (Minka)}), $p$ is chosen with the Minka criterion \cite{minka2000}, consisting in a maximisation of the training-set Bayesian evidence. This method does not require cross-validation.

\paragraph{\bf Linear shrinkage (shrinkage).} The cleaned matrix here is a convex combination of the unbiased sample estimator $E$ and a completely biased matrix, not depending on the data, that we will take as the identity matrix $\1_N$ \cite{haff1980,bouchaud2003}. Depending on $\alpha\in [0,1]$, the cleaned estimator is $C_\alpha = (1-\alpha) \1_N + \alpha E$, or $\hat\lambda_i = (1-\alpha) + \alpha \lambda_i$. In this case it is $\gamma=1-\alpha$. As explained in \cite{bun2016}, the shrinkage method corresponds, in Bayesian Random Matrix theory, to the posterior average of the covariance matrix when the prior distribution is the inverse-Wishart distribution whose mean is the identity matrix in $N$ dimensions, $\1_N$. 

\paragraph{\bf Optimal Rotationally Invariant Estimator (RIE).} The cleaned spectrum is \cite{ledoit2011,Bun_2016,bun2016,bun2016my,bun2017}:

			\begin{align}
				\hat\lambda_i=&\frac{\lambda_i}{|1-q+q z_i s(z_i)|^2} 
				\label{eq:RIE}
			\end{align}
			where $s(z):={\rm tr}[(z\1_N-E)^{-1}]/N$, $z_i:=\lambda_i-\myi \eta$, being $\myi$  the imaginary unit and $\eta$  a small parameter, coming from the limit ${\eta\to 0}$ in the derivation of the RIE estimator for large $N,T$ (through the Sokhotski–Plemelj identity, see \cite{bun2017}, sec. 4). In references \cite{bun2016,bun2017} it is explained that, for finite $N$, a convenient choice is $\eta=N^{-1/2}$. Roughly speaking, the RIE estimator is derived by imposing that matrix $C$ is, among those sharing the eigenvectors with $E$, $C=U^\dag \hat\Lambda\ U$, the one that exhibits a minimum Hilbert-Schmidt distance $d_{\rm HS}(C,\Cv)=\traza[(C-\Cv)^2]$ from the population matrix $\Cv$. Albeit the population matrix is not known, for the minimisation of $d_{\rm HS}(C,\Cv)$ it is sufficient to know, for large enough $T$, its spectral density $\rho_{\Cv}$, which is in turn related to the sample spectral density $\rho_E$ through the Marcenko-Pastur Equation (see \cite{ledoit2011,Bun_2016,bun2016,bun2017} for more details). The RIE estimator is expected to provide a better estimation, in terms of $d_{\rm HS}$, than any other algorithm modifying only the sample spectrum $\Lambda$, at least for sufficiently large values of $T$. In particular, we expect the RIE estimator to be more efficient than the PCA, shrinkage, caut-PCA and $q$-corrected raw estimators.

We have implemented two variants of the RIE algorithm, one in which the parameter $\eta$ is chosen according to the value prescribed in the literature $\eta=N^{-1/2}$ \cite{bun2016,bun2017} (simply called {\bf RIE}), the other one (called {\bf RIE (CV)}) in which $\eta$ is cross-validated on a grid of values. The RIE estimator does not require to be cross-validated (hence, for it, $\Xtr=\Xin$ and $\Tva=0$). The cross-validated hyper-parameter $\eta$ of RIE (CV) does not balance bias and variance errors, hence it does not play the role of the parameter $\gamma$ mentioned above.

			\paragraph{\bf Factor Analysis (FA).} This method proposes a cleaned matrix of the form: $($lower-rank matrix$)$ $+$ $($heteroschedastic noise diagonal matrix$)$. The generic form of the cleaned matrix is given by $C = M^\dag M + Q$. Here $M$ is a $r\times N$ real matrix, $r\in\{0,\ldots,N\}$, and $Q$ is a real diagonal square matrix of size $N$. Given $r$, the values of $M$ and $Q$ are found numerically by maximization of the inversion likelihood $\N(\Xin|C)$ (see, for example, \cite{barber2012,lillo2005}). The value of the hyper-parameter $r$ (the rank of the $M^\dag M$ matrix) is chosen by cross-validation. 

			\paragraph{\bf Graphical Lasso (Lasso).} Given the positive regularisation ($L_1$-norm) hyper-parameter $\alpha_{\rm L}$, the cleaned precision matrix $J=C^{-1}$ is given by maximization of the training likelihood minus a regularisation term \cite{friedman2007} (see also references in \cite{zhou2014}):
			\begin{align}
				J^* = \arg \max_J \left\{\ln \N(\Xtr|J^{-1}) - \alpha_{\rm L} \sum_{i<j} |J_{ij}| \right\}.
			\end{align}\\
In this case it is $\gamma=\alpha_{\rm L}$. For $\alpha_{\rm L}=0$ the estimated $C=E$, while for $\alpha_{\rm L}=\infty$, it is $C=1_N$. \\

As a comparative reference for the efficiency of the above algorithms, we propose two further, simple methods.  \\

\paragraph{\bf Cautious-PCA (caut-PCA).}  We propose the following simple variant of PCA. In this case $\gamma=N-p$ is again the number of neglected principal components, but the spectrum of the cleaned matrix is modified differently. 
\begin{enumerate}
\item First, one provisionally sets $\hat\lambda_{i\le p}=\lambda_i$ and $\hat\lambda_{i>p}=\bar\lambda_p$ as in PCA, but with $\bar \lambda_p = \lambda_p$ (the value of the lowest fitted sample eigenvalue), instead of $\bar \lambda_p = \sum_{i>p}\lambda_i/(N-p)$ as in normal PCA. 
\item {\label{sec:cauPCArescaling}} Second, the whole cleaned spectrum is hence rescaled in such a way that $C$ exhibits the same total variance as $E$: $\traza(\hat\Lambda)=\traza(\Lambda)$, or $\hat\lambda_{j}:=\kappa_p \hat\lambda_{j}$, with $\kappa_p=N/(N v_p+(N-p)\lambda_p)$ and $v_p=(\sum_{j\le p}\lambda_j)/p$.

\item Finally, the cleaned matrix is $C=U^\dag  \hat\Lambda U$. 
\end{enumerate}

Alternatively, the rescaling of the spectrum in step \ref{sec:cauPCArescaling} may be substituted by a standardisation of $C$: $C_{ij}:=C_{ij}/\sqrt{C_{ii}C_{jj}}$. In our numerical analysis, both strategies lead to almost identical results.

Both PCA and caut-PCA methods fit the $p$ largest eigenvalues and corresponding eigenvectors of $E$, and neglect the lowest $N-p$ sample eigenvalues and corresponding eigenvectors. In the $N-p$-dimensional subspace of $\mathbb{R}^N$ generated by the neglected eigenvectors, the associated cleaned matrices are degenerated with a single ``noise'' variance $\bar\lambda_p$. The difference is that in PCA the noise variance $\bar\lambda_p$ is substituted by its maximum likelihood value $\bar\lambda_p^\text{(ml)}=N(1-v_p)/(N-p)$ \cite{minka2000}, while in caut-PCA the noise variance $\bar\lambda_p^\text{(cau)}=\kappa_p\lambda_p$ consistently equals the lowest fitted eigenvalue $\lambda_p$ that is considered to be significant, apart from the normalising constant $\kappa_p$, which is close to $1$ in the relevant regime $\lambda_p\ll N v_p/(N-p)$. In this regime, and whenever $\lambda_p\ge v_p \bar\lambda_p^\text{(ml)}$, the noise variance of caut-PCA {\it is larger than its maximum likelihood value} $\bar\lambda_p^\text{(cau)}\ge \bar\lambda_p^\text{(ml)}$. In this sense, caut-PCA is more cautions. See more details in appendix \ref{sec:spectra}.

\paragraph{\bf Early-stopping Gradient Ascent algorithms (GA)} consist in an iterative updating of the correlation matrix $C$. It is initially set to $\1_N$, hence updated following a gradient ascent search of the training likelihood $\ln \N(\Xin|C)$, or $C:=C+\eta_{\rm GA}(\partial \ln \N(\Xin|C')/\partial C')|_{C}$, where $\eta_{\rm GA}$ is a constant learning rate. While the likelihood is maximized by $C=E$, we stop earlier the optimization. The stopping criterion is given by the time of first decrease of one of the validation-set criteria in Section \ref{sec:metrics}).

We have implemented and compared  mainly two variants of this algorithm, differing as follows. {\bf DGAW} (Deterministic Gradient Ascent-Wishart): the iteration is not on the matrix elements of $C$, but on those of an $N\times N$ matrix $Y$ such that $J=C^{-1}=Y Y^\dag$ (in such a Wishart form, the symmetry and positive definiteness of $C$ is guaranteed in the iterative dynamics); {\bf SGA} (Stochastic Gradient Ascent) and its Wishart form {\bf SGAW}: the iterative update algorithm for $C$ or $Y$ is not deterministic but stochastic: the gradient in each iteration $\tau$ is not that of $\ln \N(\Xin|C)$ but, similarly to mini-batch learning in machine learning \cite{mehta2019}, that of a random bootstrapping $\Xin{(\tau)}$ of the training data, different from iteration to iteration. We have as well implemented further optional variants, as the coupled dynamics of a Lagrange multiplier guaranteeing the condition $\traza(C)=\traza(E)$. Please, see further details of the GA algorithms in the appendix \ref{sec:GA}.

\subsection{Quality of the cleaned matrices according to different criteria \label{sec:metrics}}

We evaluate the quality of the cleaned matrix $C$ in the test-set according to different criteria $Q(\Xte|C)$:

\paragraph{\bf Test-likelihood (referred to as $\ell$).} The criterion $\ln {\cal N}(\Xte|C)/\Tte$ is the average of the logarithm of the Gaussian likelihood over the test-set vectors $\x(t)$: $\ln {\cal N}(\Xte|C)/\Tte=-(1/2)[ \ln (2\pi) + \ln \det C + {\rm tr} (C^{-1} E_{\rm te}) ]$.
\paragraph{\bf Test-pseudo-likelihood.} We compute the average, over the test-set vectors $\x(t)$, of the pseudo-likelihood $\ln{\cal L}(\x)=\sum_{i=1}^N \ln p_i(x_i|\x_{/i},C)$, where ${\bf x}_{/i}$ is the vector ${\bf x}$ with missing $i$-th coordinate, and where $p_i(x_i|\x_{/i},C)$ is the marginal of $\N(\x|C)$ given all the coordinates but $x_i$; it is a univariate normal distribution $p_i(x_i|\x_{/i},C)=\N(x_i-\mu_i|\sigma_i^2)$ with $C$- and $\x_{/i}$-dependent average $\mu_i$ and variance $\sigma_i^2$:
		
		\begin{align}
	&		\ln {\cal L}(\Xte|C) = \nonumber\\
 &=\frac{1}{N\Tte} \sum_{t=1}^\Tte \sum_{i=1}^N \ln \N\big(x_i(t)-\mu_i(\x(t),C) | \sigma^2_i(C)\big)  \\
	&		\mu_{i} (\x,C):= -\frac{\sum_{m\ne i} J_{im} x_m}{J_{ii}}, \qquad  \sigma_i^2(C) := {J_{ii}}^{-1} \label{eq:mu_pseudolikelihood}
		\end{align}
and where $J=C^{-1}$.

\paragraph{\bf Test-completion error (referred to as $\bar c$).} We define the {\it completion error} of the $i$-th coordinate of vector $\x$, $c_i$, as the absolute value of the difference between $x_i$ and its expected value according to the marginal distribution $p_i(\cdot|\x_{/i},C)$, or: $c_i:=|x_i-\mu_i(\x,C)|$. The {\it completion error of the dataset} $\Xte$, $\bar c(\Xte|C)$ is defined as the average of the single coordinate completion error $c_i$ over all coordinates $i$ and all vectors $\x$ in $\Xte$. It is a variant of the pseudo-likelihood in which $\sigma_i$ is not taken into account:
		
		\begin{align}
			\label{eq:completionerror}
			\bar c(\Xte|C) &= \frac{1}{N \Tte} \sum_{t=1}^\Tte \sum_{i=1}^N \big| x_i{(t)}- \mu_i(\x(t),C))\big|  
			.
		\end{align}
The completion error is, hence, interpretable: it is the error, in units of the coordinates' variance ($=1$), of the expected value of the missing coordinates $x_i$, equation \ref{eq:mu_pseudolikelihood}, according to the Gaussian model induced by the inferred $C$.

\paragraph{\bf Distance to the true precision and correlation matrices (referred to as $d$).} If the generative model of the data defined by the probability density $\Pv$ is known, it is possible to evaluate the quality of the cleaned matrix by computing the similarity between $C$ and $\Cv$ (being $\Cv_{ij}=\<x_ix_j\>_{\Pv}$) and between $J=C^{-1}$ and $\Jv=\Cv^{-1}$, according to a given criterion. We will consider the matrix-element-wise metrics:

		\begin{align}\label{eq:matrixmetrix}
			d(\Jv,J) &= \frac{\sum_{i\le j} |\Jv_{ij} - J_{ij}|}{\sum_{i\le j} |\Jv_{ij}|} 
		\end{align}
		and equivalently for $d(\Cv,C)$. This metrics is essentially equivalent to the Hilbert-Schmidt distance $d_{\rm HS}(\Cv,C)={\rm tr}[(\Cv-C)^2]$. Indeed, {\it given the generative model}, the relative efficiency of various algorithms as presented in section \ref{sec:results} is qualitatively equal using $d$ or $d_{\rm HS}$. The only essential difference is that in equation (\ref{eq:matrixmetrix}) the mean error between matrix elements is expressed in units of the mean absolute value of the  population matrix elements. The results are again essentially unchanged using a variant of $d(\cdot,\cdot)$ in which one discards the diagonal (i.e., $i<j$ in (\ref{eq:matrixmetrix})). Notice that this metrics is interpretable: $d(\Jv,J)$  is the average distance between the matrix elements of the cleaned and true precision matrices, in units of the average value of the true precision matrix elements.

\subsection{Hyper-parameter tuning \label{sec:crossval}}

Hyper-parameter tuning for the algorithms in Section \ref{sec:algorithms} is performed by $K$-fold cross-validation with $K=6$, using the four quality criteria $Q$ defined in Section \ref{sec:metrics}. In other words, the optimal hyper-parameter $\pi^*$ is chosen by maximisation of $\<Q(\Xva|C_\pi)\>$, where $C_\pi$ is computed from $\Xin$ and the average is over the $K$-fold partitions of $\Xtr=(\Xin,\Xva)$.

\subsection{Synthetic data generation \label{sec:generative}}

For the generation of the synthetic data, we employ a generative multivariate Gaussian model ${\cal N}(\cdot|\Cv)$ whose population covariance matrix is drawn from a probability distribution over correlation matrices, that we dub the Dirichlet-Haar model. In the Dirichlet-Haar model $\Cv$ is defined through the spectral decomposition $\Cv=W^\dag \tilde\Lambda W$ where $W$ is drawn from the Haar distribution (the uniform distribution over orthogonal matrices) and $\tilde\Lambda$ is a diagonal matrix whose eigenvalues $\tilde{\bm\lambda}$ are drawn from the Dirichlet distribution with parameter $\alpha_\text{D}$: $\tilde{\bm \lambda}/N\sim \text {Dir}(\cdot|\alpha_\text{D})$ so that $\traza(\Cv)=\traza(E)=N$. Therefore, $\alpha_\text{D}$ is the parameter that determines the degree of homogeneity/sparsity of the spectrum of $\Cv$: large values of $\alpha_\text{D}$ lead to homogeneous eigenvalues $\tilde\lambda_i$, hence to correlation matrices with off-diagonal elements much lower than the diagonal (for $\alpha_\text{D}\to\infty$, the Dirichlet-Haar model converges to the delta distribution around $\Cv=\1_N$). Vice-versa, lower values of $\alpha_\text{D}$ lead to larger (and $W$-dependent) off-diagonal correlation (and precision) matrix elements, with a single large eigenvalue close to $N$. For this reason, $\alpha_\text{D}$ (or, more precisely, $\alpha_\text{D}^{-1}$) may be seen as a measure of the degree of interaction strength of the population precision matrices $\Jv$.

Summarising, given $N$, $q=N/T$ and $\alpha_\text{D}$, we generate $\Ns$  synthetic datasets, representing the ``subjects'', according to the following procedure: we first sample an orthogonal matrix $W$ from the Haar ensemble, and a set of eigenvalues from the Dirichlet distribution $\y \sim \text {Dir}(\cdot|\alpha_\text{D})$, $\tilde{\bm \lambda}=N \y$; we construct $\Cv = W^\dag \tilde\Lambda W$ and sample $T$ vectors from the resulting Gaussian distribution $\x(t)\sim{\cal N}(\cdot|\Cv)$. Such vectors $(\x(1),\ldots,\x(T))$ constitute the synthetic sample $X$. We repeat the procedure $\Ns$ times, getting a $\Ns\times N\times T$ synthetic collection of datasets $(X^{(s)})_{s=1}^\Ns$, in such a way that to every subject corresponds a different correlation matrix, with different (random) eigenvectors and different (random, but with common interaction strength) eigenvalues. 

\subsection{fMRI data \label{sec:fMRIdata}}

We analyse two fMRI dataset collections of BOLD activity time series of human subjects at rest, called ``A'' and ``B''. The collection A is the one analysed and described in references \cite{mastrandrea2017,mastrandrea2021}. The collection B  is the large population-derived CamCAN Data Repository, from the Cambridge Center for Aging and Neuroscience \cite{camcan,shafto2014,TAYLOR2017}. In collection A we have measurements of $\Ns=40$ subjects, with $T=180$ observations of $N=116$ features, that we randomly split in $\Ttr=144=\Tin+\Tva=120+24$, and $\Tte=36$ for those algorithms necessitating a validation set, otherwise $\Ttr=\Tin=144$.  
For the collection B, we have $\Ns=652$, $N=114$,  and $T=260$ observations split in $\Tin=174$, $\Tva=34$, and $\Tte=52$.
\section{Results \label{sec:results}}

\subsection{Noise-cleaning the synthetic dataset\label{sec:resultssynthetic}}

We applied the methods from Section \ref{sec:algorithms} to the synthetic datasets described in Section \ref{sec:generative} with varying dimensions per sample $q=N/T$ and degree of interaction $\alpha_\text{D}^{-1}$. 
We assess the quality of the cleaned correlation and precision matrices according to the criteria described in Section \ref{sec:metrics}, evaluated on the test set of each subject.

We consider a grid of values of $q, \alpha_\text{D}$ constructed as follows: We take $N=116$ fixed, that coincides with the dimension of the standard fMRI parcelization of the human brain in $116$ regions of interest, and coinciding with value of $N$ for fMRI collection A, see Section \ref{sec:fMRIdata}. The value of $\Ttr$ takes the values $144,200,300,1000,2000$. The lower value $\Ttr=144$ coincides with $\Ttr$ of the fMRI collection A. $\alpha_\text{D}$ takes the values $\alpha_\text{D}=0.5,1,1.5,2,2.5,3,4$. 
A fraction 1/6 of the  $\Ttr$ observations is used for validation, as we said in sec. \ref{sec:crossval}. The test sets are composed by $\Tte=0.25\,\Ttr$ vectors. 
For each value of the couple $q, \alpha_\text{D}$ the results are averaged over $\Ns=10^2$ realizations of the dataset (different ``subjects'').

\begin{figure}[h]
\includegraphics[width=0.9\columnwidth]{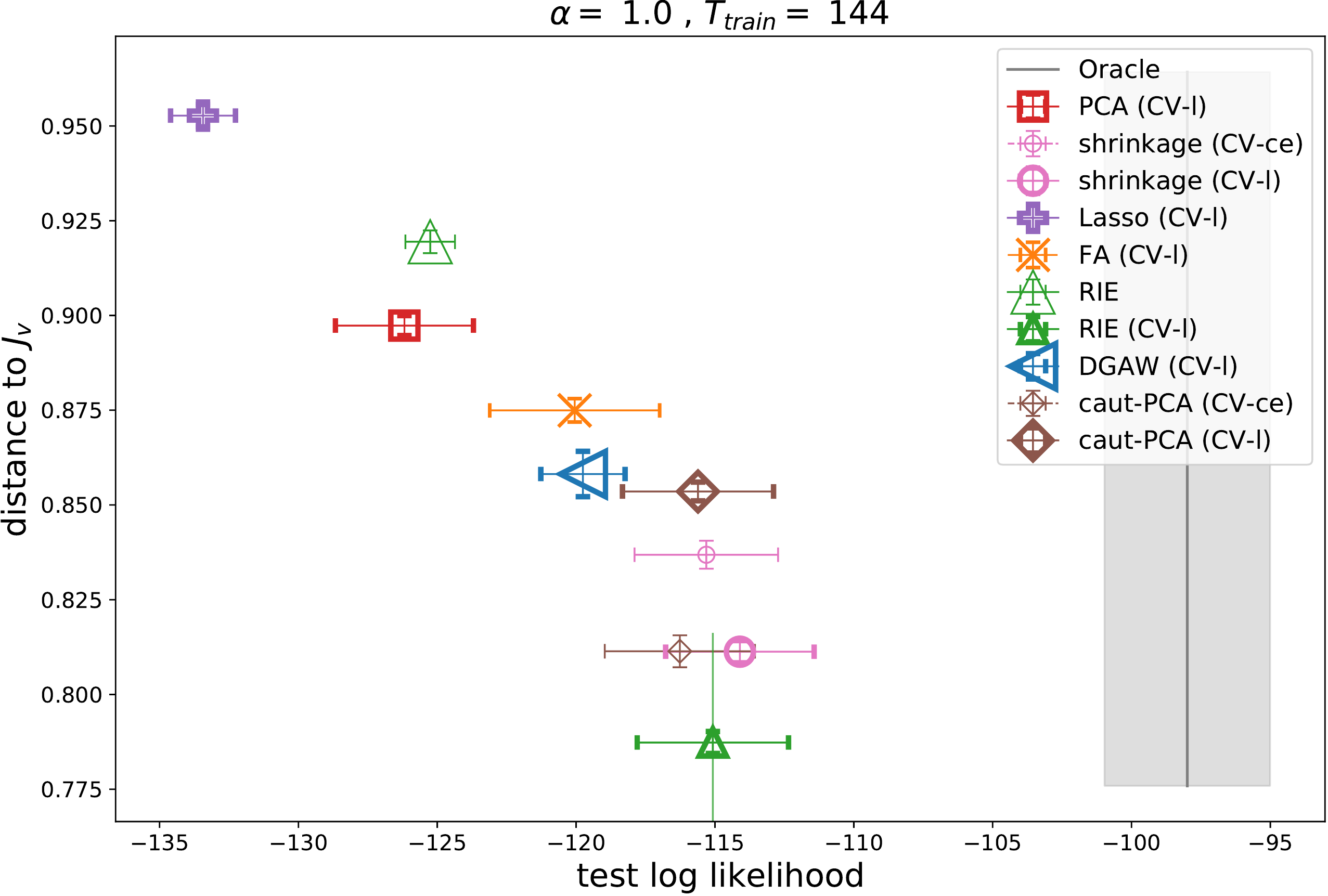} \\
\includegraphics[width=0.9\columnwidth]{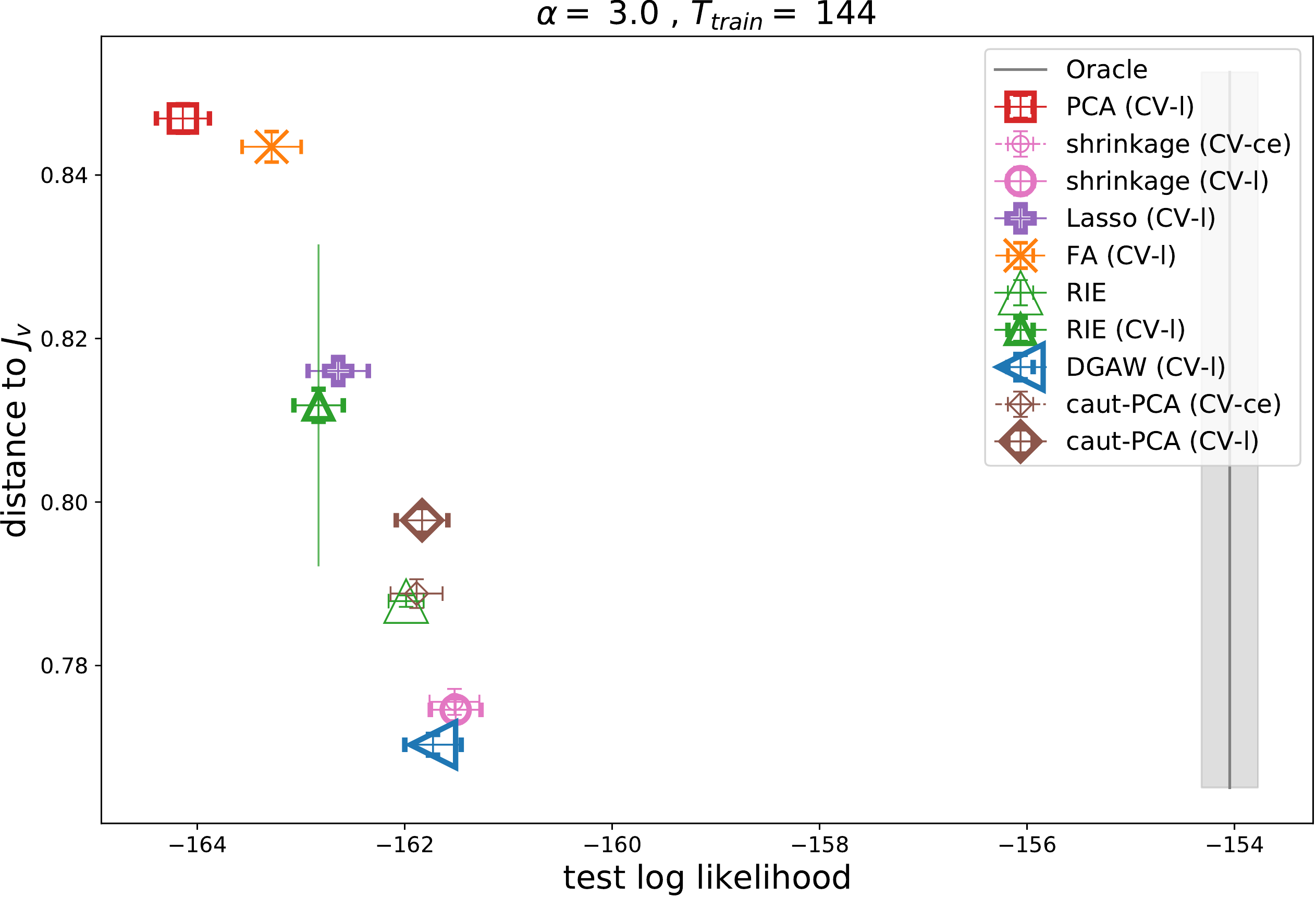} \\
	\caption{ \label{fig:observable_scatter_dirichlet1} Scatter plot of $d(\Jv,J)$ (lower is better) versus $\ell$ (higher is better) for synthetic data in the severe undersampling regime $\Ttr=144$. Points and error bars are averages and standard errors of the mean across subjects, and each point corresponds to a different method. The thin vertical error bars with no cap over RIE are the standard deviation across subjects. The vertical line indicates the likelihood of the Oracle method (whose $d(\Jv,J)$ vanishes). Higher panel: $\alpha_\text{D}=1$ (strongly correlated matrices, highly discontinuous spectrum). Lower panel: $\alpha_\text{D}=3$ (weakly correlated matrices).}
\end{figure}

In this section and in the appendix \ref{sec:systematic} we present results for the methods: {\bf PCA (CV-l)}, {\bf PCA (Minka)},  {\bf shrinkage (CV-l)}, {\bf Lasso (CV-l)}, {\bf FA (CV-l)}, {\bf RIE}, {\bf RIE (CV-l)}, {\bf RIE (CV-e)}, {\bf DGAW (CV-l)}, {\bf SGAW (CV-l)}, {\bf SGAW (CV-e)}, {\bf caut-PCA (CV-l)}, {\bf caut-PCA (CV-e)}. The first part of the methods' name is as explained in \ref{sec:methods}, while the notation CV-l and CV-e refers to the cross validation strategy: by maximisation of the test-likelihood or test-completion error, respectively (see \ref{sec:crossval}). For the sake of clearness, in the figures of this section we omit the combinations of methods and cross-validation strategies that are less distinguishable or efficient. Although the cross-validation strategy may induce some statistically significant differences for some values of the parameters $q,\alpha_\text{D}$ (see figure \ref{fig:observable_scatter_dirichlet1}, $\alpha_\text{D}=1$, $\Ttr=144$), such differences are negligible in most cases. Importantly, in the  analysis below we add the  baselines {\bf Oracle} and {\bf raw} in which no cleaning procedure is applied but, instead, the quality estimators $Q(\Xte|\Cv)$ and $Q(\Xte|E)$ are directly evaluated using as ``cleaned'' matrices the true and the sample matrices, respectively. An efficient estimator is expected to exhibit a value of $Q$ larger than that of the raw estimator, and as close as possible to that of the Oracle estimator. 

We mainly focus on the analysis of the distance from the true precision matrix $d(\Jv,J)$. This is the quality criterion presenting by far higher variability across methods in terms of its subject-to-subject errors (see figure \ref{fig:observable_scatter_dirichlet1}). Information regarding the performance in terms of other quality criteria (test-likelihood, $d(\Cv,C)$, test-completion error, test-pseudo-likelihood) and for variants of these cleaning methods (cross-validated with respect to the completion error or the pseudo-likelihood) may be found in appendix \ref{sec:systematic} and in the repository \cite{repository}, where a complete table of the performance of all algorithms according to all the criteria is made available \footnote{See the python-Jupyter notebook \textsf{plot\_estimators.ipynb} in reference \cite{repository}}. 

\begin{figure}[h]
\includegraphics[width=0.9\columnwidth]{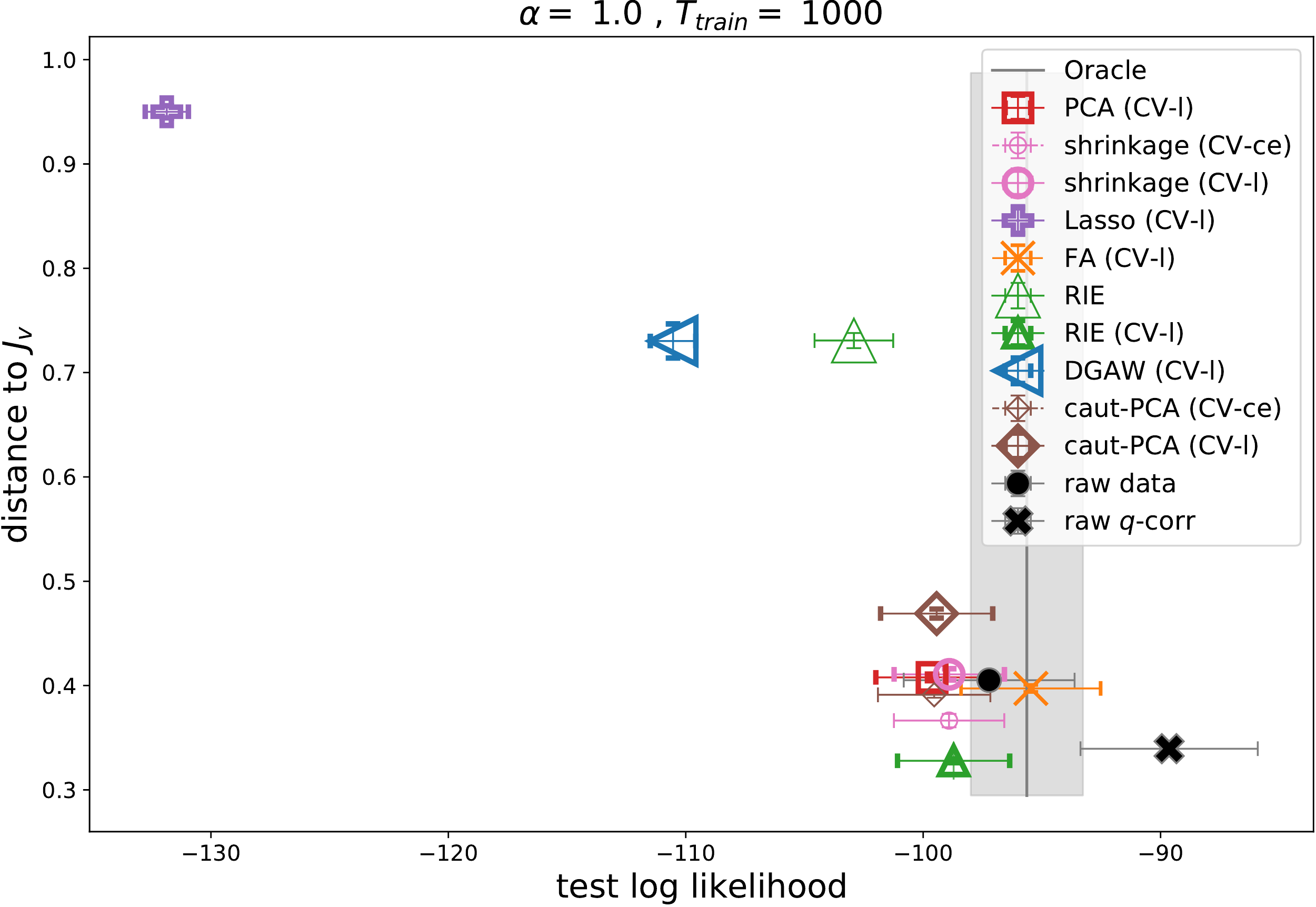} \\
\includegraphics[width=0.9\columnwidth]{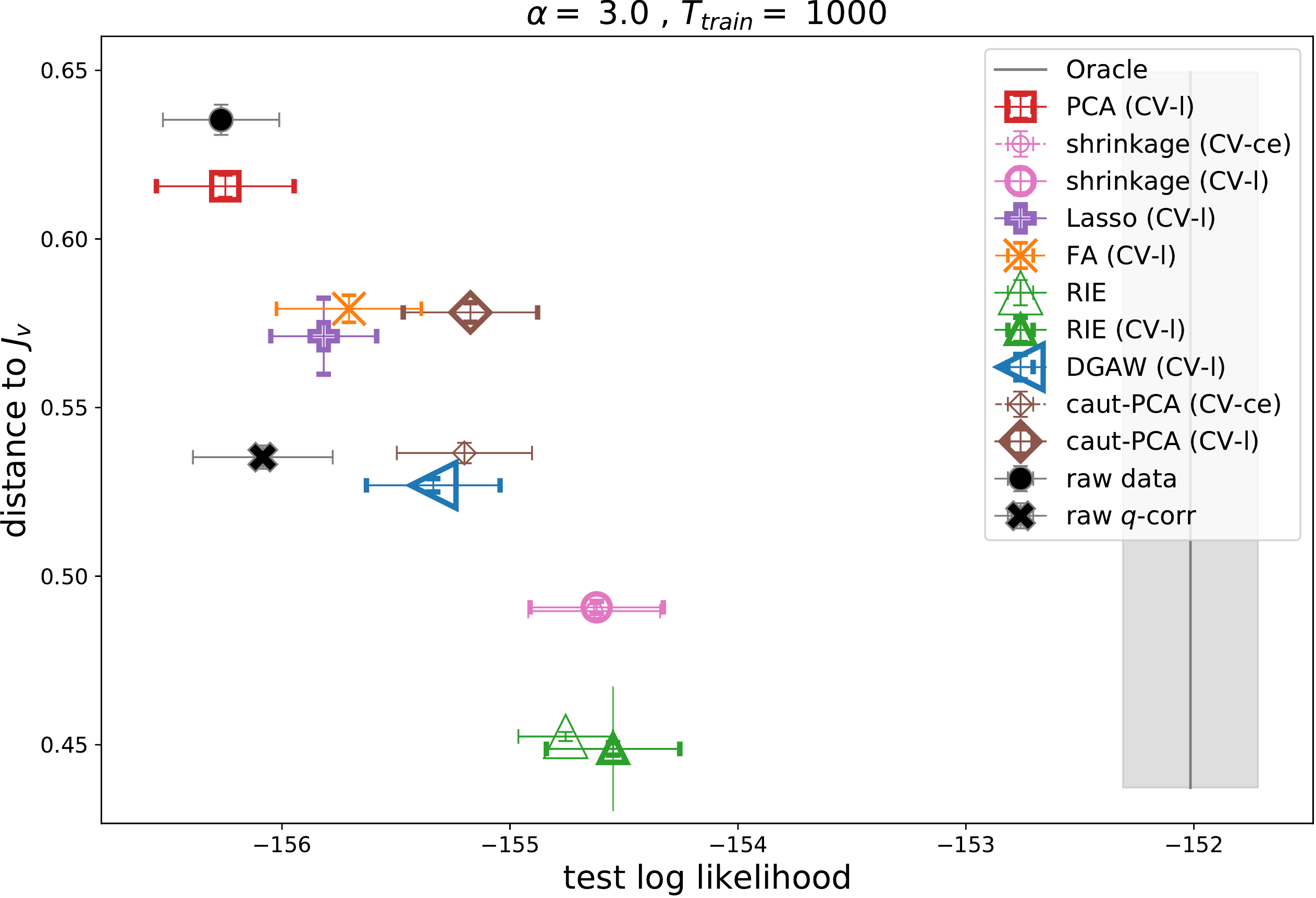} 
	\caption{ \label{fig:observable_scatter_dirichlet2} As in figure \ref{fig:observable_scatter_dirichlet1} but for $\Ttr=1000$ (moderately undersampled regime).}
\end{figure}

In figures \ref{fig:observable_scatter_dirichlet1},\ref{fig:observable_scatter_dirichlet2} we show the subject-averaged values of $d(\Jv,J)$ (being $J=C^{-1}$ the best precision matrix for a given method) versus the test-likelihood. Each point corresponds to a different algorithm, and each figure to a different combination of the dataset parameters  $q,\alpha_\text{D}$. The error bars represent the Standard Error of the Mean (SEM) across subjects. For reference, we also include the standard deviation across subjects ($\Ns^{1/2}=10$ times larger than the SEM) {\it only for the RIE (CV-l) method} ($y$-axis thin error bars without cap). Please notice that the SEM, indicated by capped error bars are, instead, of the same order than the symbol size. The grey vertical strip (Oracle method) indicates the value of the test-likelihood error corresponding to the true correlation matrix $\Cv$. Mind that the Oracle estimator exhibits null $d(\Jv,J=\Jv)=0$. 

Figure \ref{fig:observable_scatter_dirichlet1} corresponds to $\Ttr=144$ ($q\simeq 0.85$, severe undersampling), while figure \ref{fig:observable_scatter_dirichlet2}, to $\Ttr=1000$ ($q=0.116$, moderate undersampling). In the severe undersampling regime, all the considered methods lead to a cleaned precision matrix whose error is lower than one, while the distance $d(\Jv,J=E^{-1})$ corresponding to the raw estimator lies far outside the figure, at $(\ell,\bar c,d)\simeq(-327,0.79,11.7)$ for $\alpha_\text{D}=1$ and $(\ell,\bar c,d)\simeq(-381,1.5,17.6)$ for $\alpha_\text{D}=3$. In such a severe undersampling situation, the raw unbiased estimator of the precision matrix exhibits matrix element-wise errors which are more than ten times larger than the average of the matrix elements of $\Jv$. 

This is precisely what we expect from Random Matrix Theory: whatever distribution the $\Cv$ has been sampled from, the empirical matrix $E$, follows (assuming a Gaussian data likelihood) the Wishart distribution $P_\text{W}(E|\Cv)$ whose average is $\Cv$ (see, for example, \cite{bun2017}), while the sample precision matrix $E^{-1}$ follows the inverse-Wishart distribution $P_\text{iW}(E^{-1}|\Cv^{-1})$ whose average is $(1-q)^{-1}\Cv^{-1}$. Hence, we expect that the precision matrix in a situation of $q\simeq 0.9$ is about ten times larger than the true precision matrix. This argument suggests to compare our results with an additional, reference cleaning method simply consisting in multiplying the raw precision by $1-q$:

\begin{align}
E\to J = (1-q) E^{-1}
.
\label{eq:qcorrection}
\end{align}
We refer to this method to infer $J$ as {\bf raw ($q$-corr.)} in the figures legend. The raw $q$-corrected method systematically reduces the distance to the true precision matrix with respect to $E^{-1}$ ($d\simeq 2.0$ for $\alpha_\text{D}=1$ and $d\simeq 3.3$ for $\alpha_\text{D}=3$) for the lowest $\Ttr$ but, in this case, it still leads to a much larger distance $d$ than the rest of the considered cleaning methods. The situation is the opposite for the largest $\Ttr=1000$, see figure \ref{fig:observable_scatter_dirichlet2}: for $\alpha_\text{D}=1$ there is no cleaning method leading to a significantly lower $d$ than the raw $q$-corrected method: cleaning is likely to be counter-productive (only RIE (CV) leads to a non-larger $d$ than raw ($q$-corr.)). The intermediate situation, for $\Ttr=300$ is shown in figure \ref{fig:observable_scatter_T300}.

{\bf We now draw some conclusions on the synthetic data analysis} from the results in figures \ref{fig:observable_scatter_dirichlet1},\ref{fig:observable_scatter_dirichlet2} and appendix \ref{sec:systematic}.\\  
{\bf (\ref{sec:resultssynthetic}-1)} The across-method differences in terms of distance from the true precision matrix are more significant than in terms of test-likelihood and completion error: significant differences between two methods in $\ell$ and $\bar c$ also imply significant differences in $d$, while the opposite does not hold (see figures \ref{fig:observable_scatter_dirichlet1},\ref{fig:observable_scatter_dirichlet2}). In any case, the method ranking resulting from $d$, $\bar c$ and $\ell$ are consistent, these quantities being strongly correlated across methods (figures \ref{fig:observable_scatter_T144},\ref{fig:observable_scatter_T300}).  

The across-method differences in $d$ are, in some cases, significant not only in terms of SEM but even in terms of standard deviation across subjects (figure \ref{fig:observable_scatter_dirichlet1}). In these cases, the difference between the best and the worst algorithms' average $d$ amounts to two or more standard deviations of $d$ across subjects and, consequently, the best methods present lower distance than the worst methods {\it for most of the subjects}.\\ 
{\bf (\ref{sec:resultssynthetic}-2)} The ranking of methods providing a lower distance to the population precision matrix depends much on the dataset characteristics $q,\alpha_\text{D}$. 
While for high values of $q$ all the considered noise-cleaning algorithms reduce the distance to the population precision matrix, beyond the raw and raw ($q$-corr.) algorithms, for low values of $q$ the noise-cleaning may be counter-productive (figure \ref{fig:diffJtrue}). \\
{\bf (\ref{sec:resultssynthetic}-3)} For sufficiently large values of $\Ttr$, the Optimal Rotationally Invariant Estimator is the best performing in terms of all the criteria, {\it  when complemented with the cross-validation for the parameter $\eta$} suggested in this paper (algorithm {\bf RIE (CV)}, see figures \ref{fig:observable_scatter_dirichlet2},\ref{fig:diffJtrue}). \\
{\bf (\ref{sec:resultssynthetic}-4)} Considering the whole grid of dataset parameters $q,\alpha_\text{D}$, the best performing algorithms (according to $d$) are: RIE (CV), shrinkage (CV), DGAW (CV-l) (see figures \ref{fig:observable_scatter_dirichlet1},\ref{fig:diffJtrue}). The first two, however, have the advantage of being principled, faster (not requiring an optimisation at the level of $\Xin$) and robust (performing well in all regimes of the synthetic benchmark and on the fMRI data as well -- see below), while GAW works well for low degree of interaction only. The high-$\alpha_\text{D}$, high $q$ regime in which the GAW algorithm performs well is, incidentally, the most difficult regime, presenting the highest values of $d$, $\bar c$ and the lowest values of $\ell$ (figures \ref{fig:diffJtrue},\ref{fig:observable_scatter_T300},\ref{fig:observable_scatter_T144}). The results of this section (limited to the specific Haar-Dirichlet generative model that we use to generate the synthetic data) suggest to use the RIE (CV) algorithm, since {\it it is the one providing a distance $d$ lower or statistically compatible with the ($q$-corrected) raw estimator even for large values of $\Ttr$}. \\
{\bf (\ref{sec:resultssynthetic}-5)} The proposed algorithm {\bf caut-PCA} significantly improves PCA for almost all considered values of $q,\alpha_\text{D}$.

We make notice that, for some of the probed values of $q,\alpha_\text{D}$, RIE (CV) performs worst than RIE (figure \ref{fig:diffJtrue}), despite the fact that the parameter $\eta$ is cross-validated from a list that actually contains the value $\eta=N^{-1/2}$ used by the plain RIE method. This is possible since RIE does not have any hyper-parameter and consequently does not need a validation set: the spectrum $\lambda$ in equation (\ref{eq:RIE}) is computed from the $\Ttr$ vectors in $\Xtr$, while in RIE (CV) it is computed from the $\Tva=(5/6)\Ttr$ vectors in $\Xva$. The same happens with PCA (CV) and PCA (Minka) (figure \ref{fig:diffJtrue}).

\subsection{Noise-cleaning the fMRI brain activity datasets \label{sec:resultsfMRI}}

We have applied the noise-cleaning algorithms to the two fMRI collections A and B described in \ref{sec:fMRIdata}. In this case, we do not have a population precision matrix $\Jv$ to compute the distance $d$ from the inferred $J$. We assess the quality in terms of the criteria $\ell,\bar c$. 

\begin{figure}[h]
\includegraphics[width=.9\columnwidth]{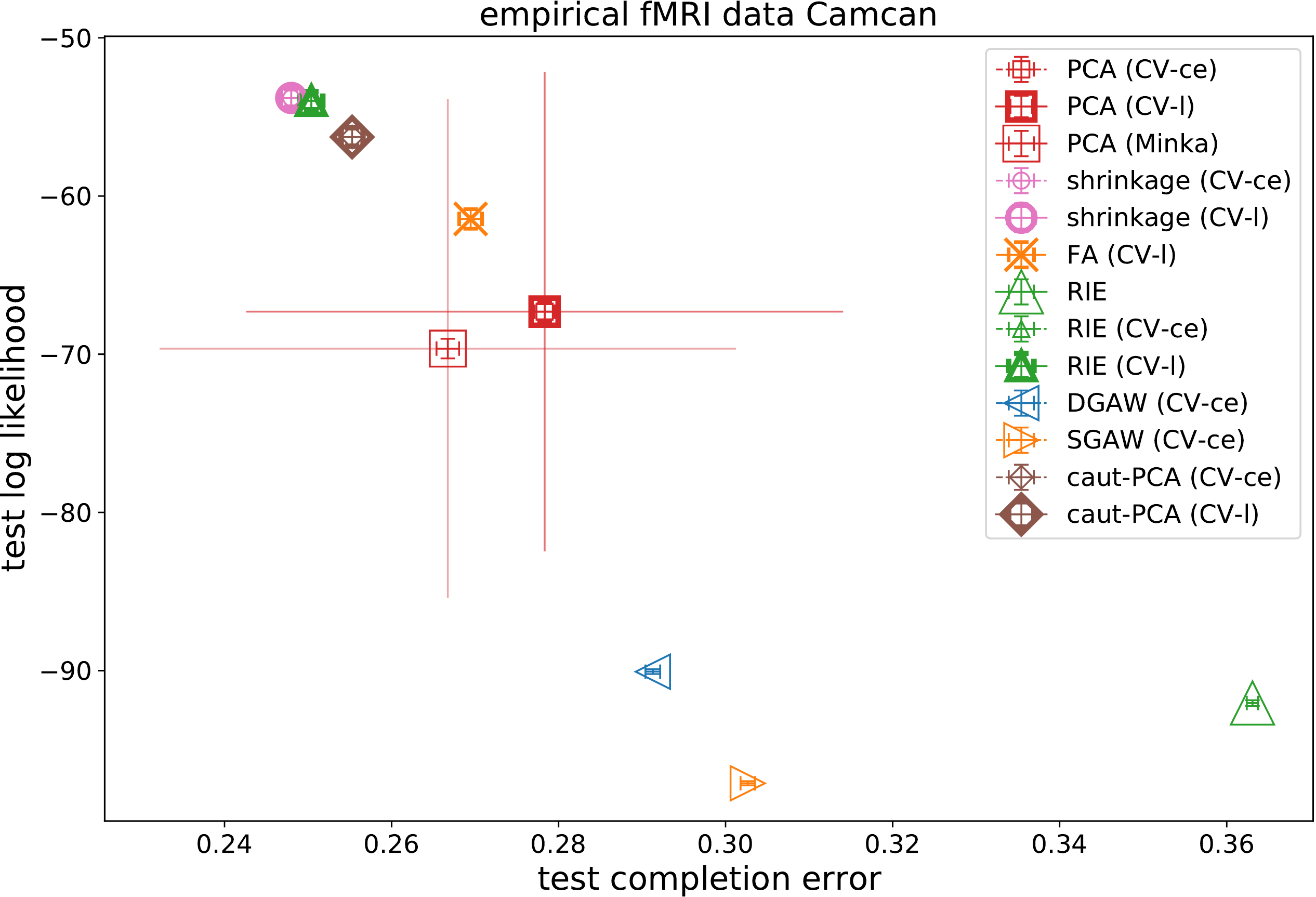} 
\includegraphics[width=.9\columnwidth]{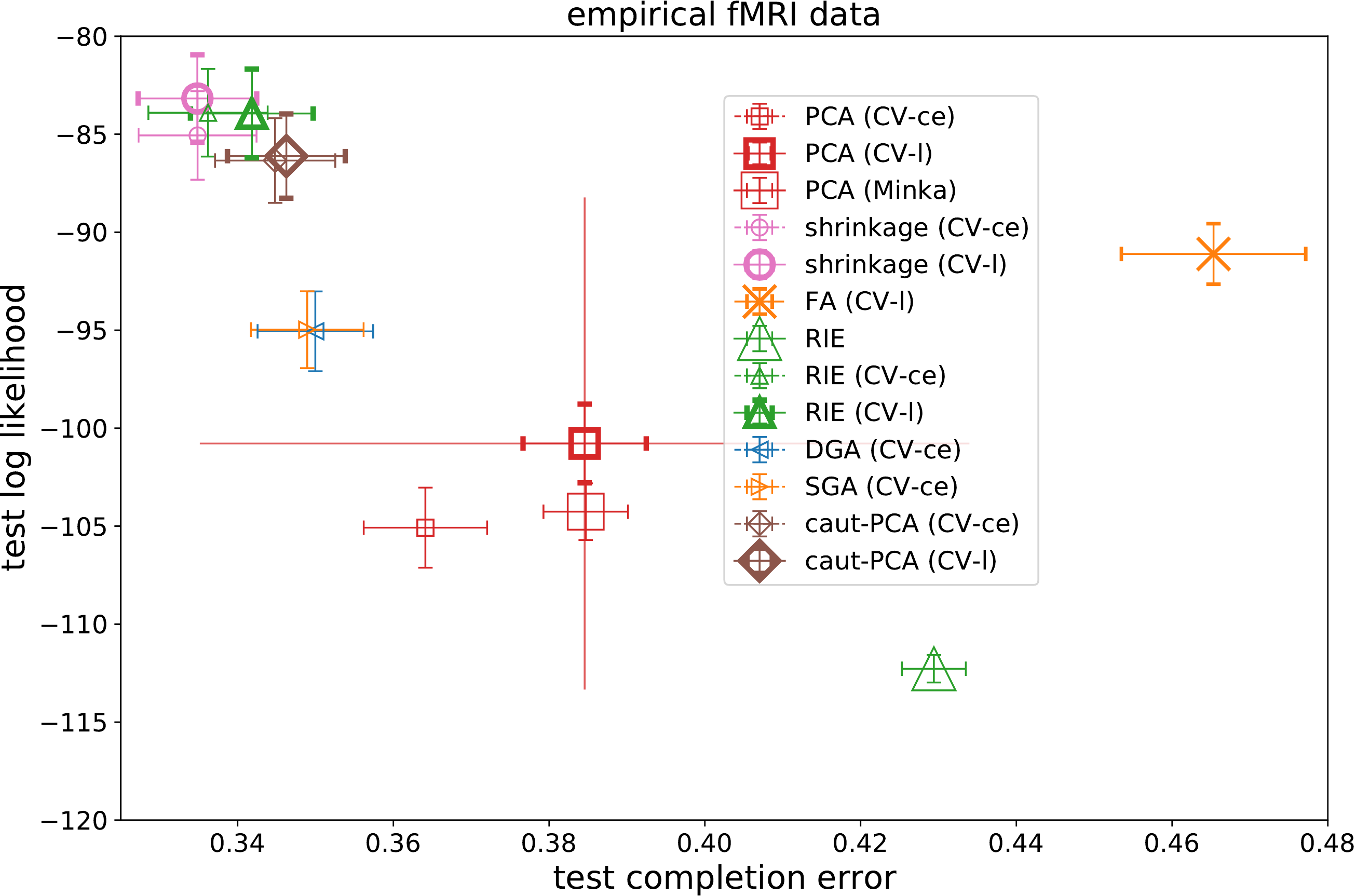} 
	\caption{ \label{fig:observable_scatter_fMRI}  Scatter plot of test-set likelihood $\ell$ (higher is better) versus completion error $\bar c$ (lower is better) in empirical fMRI data. Points and error bars are averages and standard errors of the mean across subjects, and each point corresponds to a different method. The thin vertical error bars with no cap over PCA are the standard deviation across subjects. Higher and lower panels: fMRI databases A and B, respectively.}
\end{figure}

We present, in figure \ref{fig:observable_scatter_fMRI}, the average  of the criteria $\ell,\bar c$ across the subjects of collections A and B for various noise-cleaning methods. The short, capped error bars indicate the SEM across dataset subjects, while the thin error bars without cap over the PCA (CV-l) method indicate the standard deviation. The Lasso (CV-l,e) algorithms are absent in the figure since they did not achieve convergence (see the details in appendix \ref{sec:details}). The GA algorithms not in the Wishart form may present problems with the positive-definiteness of the optimising matrix $C$, depending on the choice of the learning rate, this is why they are absent in collection B. Conversely, in collection A, the DGAW and SGAW algorithms have been excluded since they present worst performance (generally speaking, the performance of the gradient ascent-based algorithm is rather sensible to the choice of the learning rate and bootstrapping fraction, see appendix \ref{sec:GA}).

{\bf We draw the following conclusions for this section.} \\
{\bf (\ref{sec:resultsfMRI}-1)} The differences across algorithms are definitely statistically significant in both datasets, in some cases even at the subject level. Assuming that the results on synthetic datasets hold in this context, we expect that the across-algorithm differences in the (inaccessible) $d$ are even more significant.\\ 
{\bf (\ref{sec:resultsfMRI}-2)} The results are qualitatively consistent in the two data collections. The best performing algorithms in terms of both $\ell$ and $\bar c$ are RIE (CV) and shrinkage (CV), followed by caut-PCA, consistently with the synthetic dataset results. \\
{\bf (\ref{sec:resultsfMRI}-3)} On the natural datasets, only RIE (CV), not RIE, performs well: the cross validation of the parameter $\eta$ becomes particularly useful. We make notice that the RIE algorithm equation (\ref{eq:RIE}) is a function of $q$ and that, in natural data, the choice $q=N/\Ttr$ neglects the temporal correlation between dataset vectors. A more convenient choice would be $q_\text{eff}=N\tau/T$ where $\tau$ is the correlation time of the series (so that the effective number of uncorrelated vectors is taken to be $T_\text{eff}=T/\tau$). It is possible that the cross-validation of $\eta$ in RIE (CV) compensates for the misleading choice $q=N/\Ttr$. This is a possible origin of the inadequacy of plain RIE for natural data. \\

\section{Conclusions and discussion \label{sec:conclusions}}

The precision matrix between different brain regions, inferred from fMRI of MEG, is a fundamental quantity in the context of network neuroscience. It is widely studied as a model of structural connectivity between brain areas in the harmonic approximation, and to capture significant inter-subject and inter-group differences, beyond those encoded in the correlation matrix \cite{deligianni2011,liegeois2020,pervaiz2020,smith2013functional,dadi2019,chung2021,rahim2019,ryali2012,smith2013functional,brier2015}. 

This motivates the interest in an assessment of the absolute and relative utility of various noise-cleaning strategies for an accurate inference of the precision matrix in the context of network neuroscience. In particular, we are interested in assessing, and comparing with known methods, the efficiency of an over-fitting mitigation strategy based on Random Matrix Theory, the Ledoit-Péché or Optimal Rotationally Invariant estimator \cite{ledoit2011,bun2016} (see as well \cite{drogosz2015}), whose efficiency and potential utility has not yet been, to the best of our knowledge, addressed in the context of neuroscience. 

In this article, we have performed a numerical analysis of the relative efficiency of several well-known strategies of regularisation of the covariance (and hence precision) matrix of datasets in the $T\gtrsim N$ regime, being $N$, $T$ of the order of typical fMRI and MEG neural data. For such a comparison, we have used both synthetic datasets of Gaussian vectors, of varying inverse sample ratio $q=N/T$ and degree of off-diagonal correlation, and  two datasets of  human brain activity at rest, measured by fMRI. We have performed such a comparison in terms of both the distance $d$ between the noised-cleaned (inferred) and population (``true'') precision matrices, in the case of the synthetic datasets, and of the out-of-sample likelihood $\ell$. 

We have observed that: 
\begin{enumerate}
\item At least in our synthetic dataset, the distance $d$, or the average error in the inferred precision matrix elements, may significantly depend on the chosen cleaning strategy (that may induce typical differences of the $20\%$ in $d$ or larger). Such inter-algorithm differences in $d$ are larger than those in $\ell$. This suggests that, in a context in which the precision matrix should be inferred accurately (e.g., for classification purposes), the choice of the noise-cleaning method may be crucial.
\item The analysis of both fMRI datasets consistently suggests that the algorithms RIE (CV) and shrinkage (CV) are those providing a higher $\ell$ and, consequently, a more faithful precision matrix. The method RIE (CV) has the further advantage of being, by construction, optimal with respect to other rotationally invariant methods (as shrinkage) for large enough values of $T$, as confirmed by the synthetic dataset analysis. Indeed, 
\item In synthetic data, RIE (CV) exhibits, as expected, significantly lower $d$ for large $T$'s, being the only method that improves the raw estimator $E^{-1}$ (with the ``Marcenko-Pastur $q$-correction'' (\ref{eq:qcorrection})) in all the simulated regimes (see figure \ref{fig:diffJtrue}). 
\item The simple GA algorithms, consisting in a (train-dataset likelihood) gradient ascent iterative updating of the covariance matrix, combined with an early stopping criterion to prevent over-fitting, turn out to perform as well as the best performing algorithm for low $T$ and low correlation strength in synthetic data (figures \ref{fig:observable_scatter_dirichlet1},\ref{fig:diffJtrue}). It is not our aim to present a systematic nor rigorous study of the efficiency of such algorithms, that could be optimised in several ways (bootstrapping strategy and fraction, learning rate, initial condition, stopping criterion). We rather show numerically that, as a proof of principle, such a simple early stopping gradient ascent technique is enough to accurately infer weak correlations of strongly undersampled data, at least in the synthetic dataset at hand.
\item The {\it cautious PCA} algorithm, simply consisting in raising the value $\bar\lambda$ of the noise eigenvalues in the PCA method, systematically improves the inferred precision matrix with respect to PCA in the natural and synthetic datasets (see figures \ref{fig:diffJtrue},\ref{fig:observable_scatter_fMRI} and appendix \ref{sec:spectra}). 
\end{enumerate}

Summarising, the present analysis results suggest that, whenever accurate statistical estimators of the precision matrices are needed in brain connectivity studies, the Optimal Rotationally Invariant estimator, {\it if completed with the simple cross-validation strategy for the parameter $\eta$ proposed in this article}, is the best one in terms of robustness, accuracy and computational cost. 

In this study, we have cast the inference of brain structure from single-subject temporal fMRI of MEG series as a problem of covariance matrix noise-cleaning, hence deliberately restricting the analysis to [1] {\bf linear inference:} the data non-linearities are neglected; [2] {\bf non-causal inference:} we neglect the data temporal correlations; [3] {\bf inference from single-subject data only:} we do not exploit group information. In this precise context, we have performed a systematic comparison between well known noise-cleaning algorithms, together with a further method (RIE) based on Random Matrix Theory. 
Sequentially accounting for non-linearities and temporal correlations (\cite{morone2017,fortel2019,watanabe2013,fortel2022,niu2019,abeyasinghe2015,kadirvelu2017,hahn2019,das2014,deco2012}), and group information \cite{mejia2016,brier2015,deligianni2014,pervaiz2020,liegeois2020,varoquaux2010brain,chung2021,rahim2019} would allow addressing the relative importance of these elements in the inference of functional data. Particularly interesting could be the comparison with one of the standard tools for inferring brain structure, Dynamic Causal Modelling \cite{frassle2018generative,stephan2008}, accounting for both non-linearities and temporal correlations. We suggest as well a comparison with the recent promising algorithm \cite{drogosz2015}, rooted as well in Random Matrix Theory.

We publicly release the algorithm's implementations and the code for reproducing the experiments \cite{repository}.


\section{Acknowledgments}

We acknowledge the advise of Mike X Cohen, Fabrizio Lillo and Andrea Pagnani. Thanks to Zbigniew Drogosz, Maciej Nowak and Pawel O\'swiecimka for further discussions and for making us notice reference \cite{drogosz2015}. M. I.-B. is supported by the grant EU FESR-FSE PON {\it Ricerca e Innovazione} 2014-2020 BraVi, awarded to Stefano Panzeri.

\appendix

\bibliography{biblio_regularization.bib}

\section{Cleaned spectra \label{sec:spectra}}

\begin{figure}[h]
\includegraphics[width=.9\columnwidth]{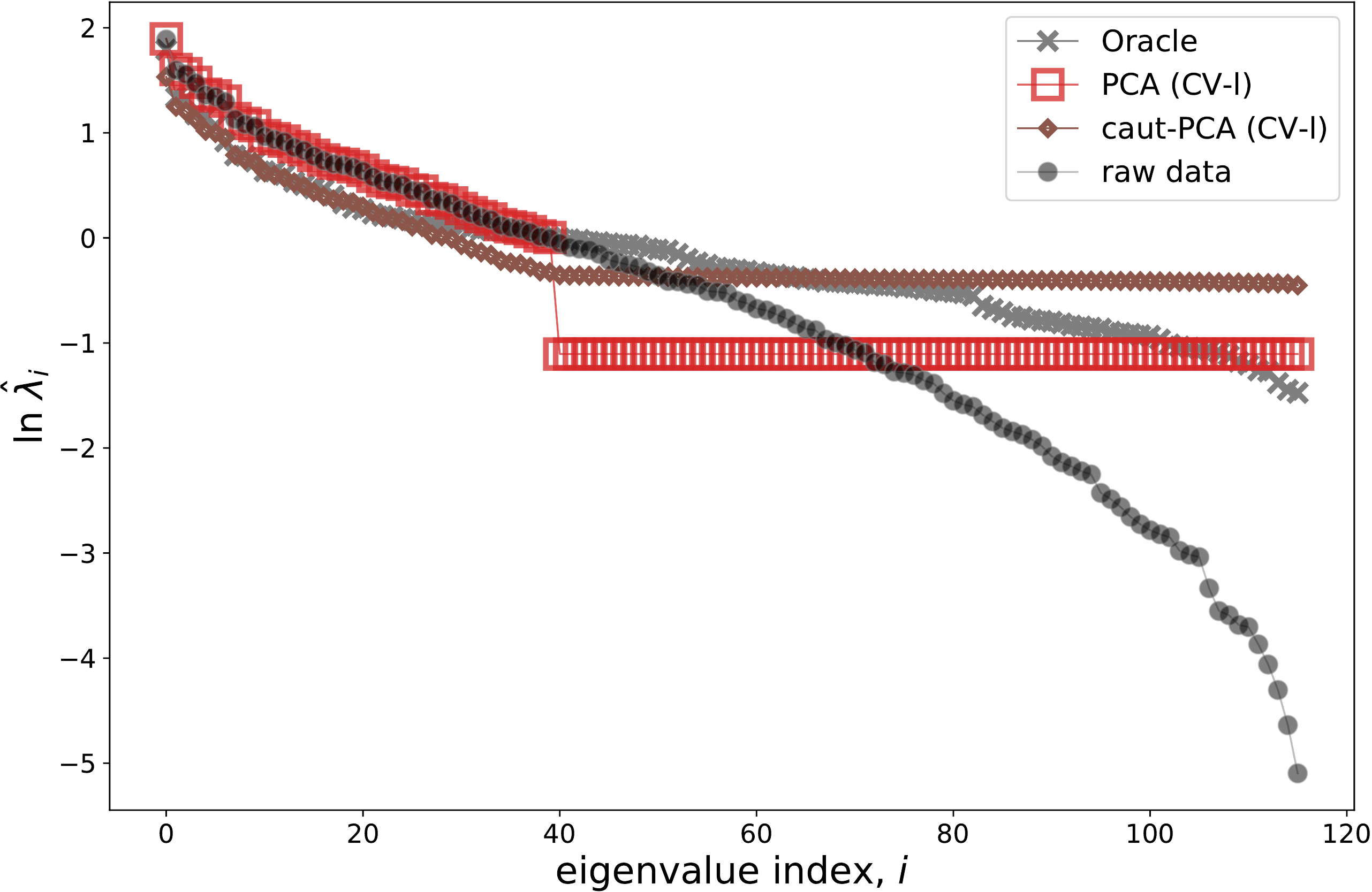} 
	\caption{ \label{fig:spectra_PCA} Logarithm of the spectra of true (Oracle), sample (raw data) and cleaned matrices (according to PCA and caut-PCA with fixed $p=40$), versus the eigenvalue index. See the details of the database in the main text.}
\end{figure}

In figure \ref{fig:spectra_PCA} we illustrate the effect of the algorithm caut-PCA. We generate a single synthetic database $X\sim{\cal N}(\cdot|\Cv)$, where $\Cv$ is sampled from the Haar-Dirichlet model with $\Ttr=144$, $N=116$, $\alpha_\text{D}=3$. Afterwards, we plot the spectra of the cleaned matrix $C$ according to PCA and caut-PCA for a fixed value of $p=40$. As explained in section \ref{sec:algorithms}, the noise eigenvalue of caut-PCA ($\bar\lambda_p^\text{(cau)}$) is larger than that of PCA ($\bar\lambda_p^\text{(ml)}$) for a fixed $p$, whenever $\bar\lambda v_p\le \lambda_p$. 

The reader may notice that the $\lambda_{j>p}$ eigenvalues of caut-PCA in figure \ref{fig:spectra_PCA} are not constant. This is because we are using the standardisation in step \ref{sec:cauPCArescaling} of the description of Cautious-PCA in section \ref{sec:algorithms}. Using the rescaling instead, one obtains a similar spectrum with a constant noise eigenvalue.

\begin{figure}[h]
\includegraphics[width=.9\columnwidth]{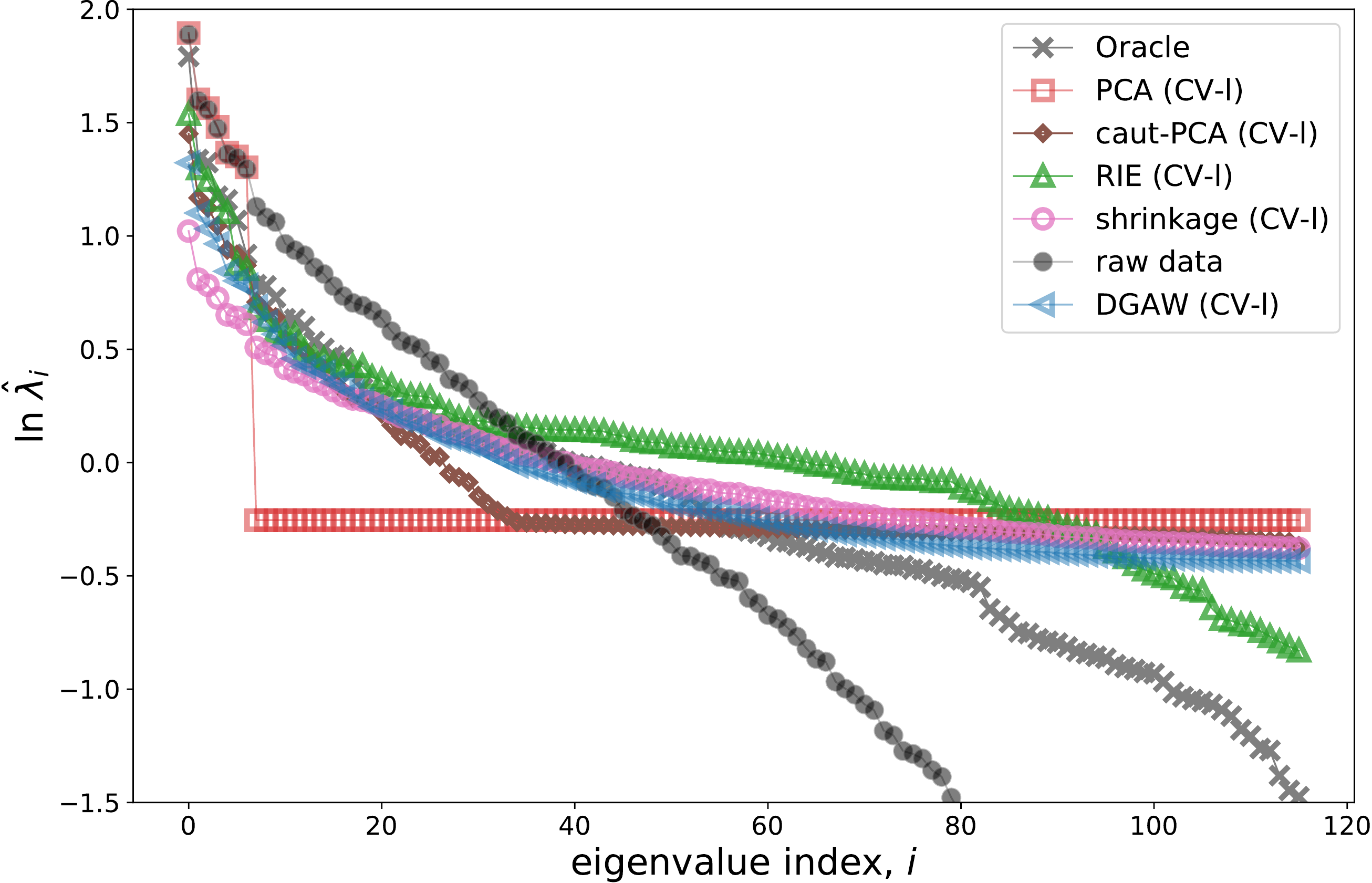} 
	\caption{ \label{fig:spectra_methods} As in figure \ref{fig:spectra_PCA}, but cross-validating the value of $p$ and comparing (in the same database $X$) PCA (CV) and caut-PCA (CV) with other methods.}
\end{figure}

As a consequence of $\bar\lambda_p^\text{(cau)} > \bar\lambda_p^\text{(ml)}$, the cross-validated value of $p^*$ tends to be lower in PCA (CV) than in caut-PCA (CV), as illustrated in figure \ref{fig:spectra_methods}. The reason is that the low value of $\bar\lambda_p^\text{(ml)}$ penalises large values of $p$. In the presence of over-fitting, for low $T$, the validation-set energy term in the likelihood, $-(1/2)\sum_t\sum_{j>p}(x_j'(t))^2/\bar\lambda_p$ (being $\x'=U\x$) decreases fast with $p$, since the average of $(x'_{j>p})^2$ over the validation-set tends to be larger than in the inversion-set for low $T$, i.e., larger than its maximum likelihood value $\bar\lambda^\text{(ml)}_p$ (as predicted by the Marchenko-Pastur equation). Raising the value of $\bar\lambda_p>\bar\lambda^\text{(ml)}_p$, one takes into account this fact. Therefore, the resulting value of the cross-validated $p^*$ tends to be larger in caut-PCA. As a consequence (see figure \ref{fig:spectra_methods}), a larger number $p^*$ of eigenvalues is more similar to the sample (and, more importantly, to the Oracle) spectrum. 

For reference, in figure \ref{fig:spectra_methods} we also compare the methods PCA (CV) and caut-PCA (CV) with: shrinkage (CV), RIE (CV), GAW, raw, and Oracle. 

\section{Systematic results for the synthetic dataset \label{sec:systematic}}

\begin{figure}[h]
\includegraphics[width=.9\columnwidth]{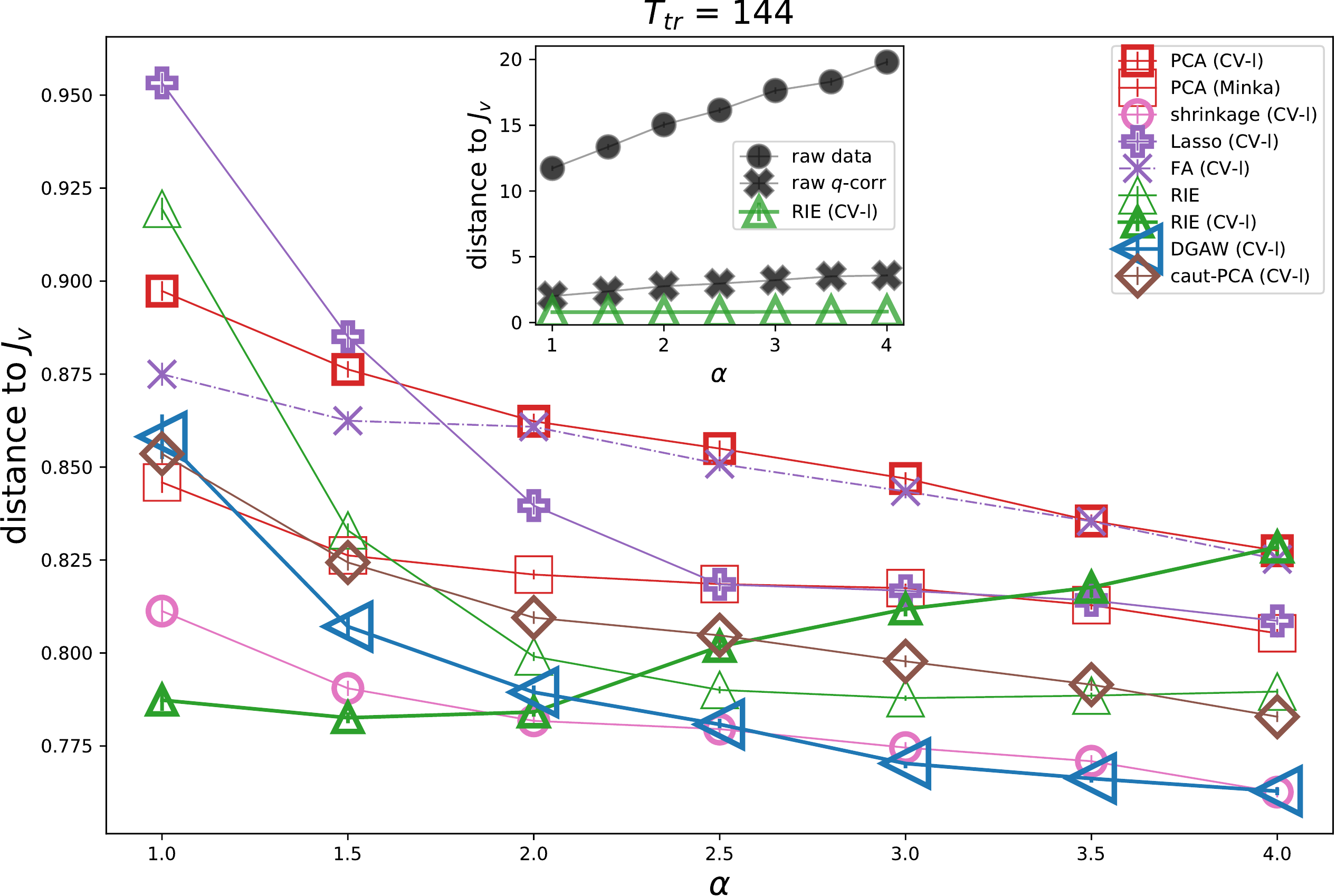} \\
\includegraphics[width=.9\columnwidth]{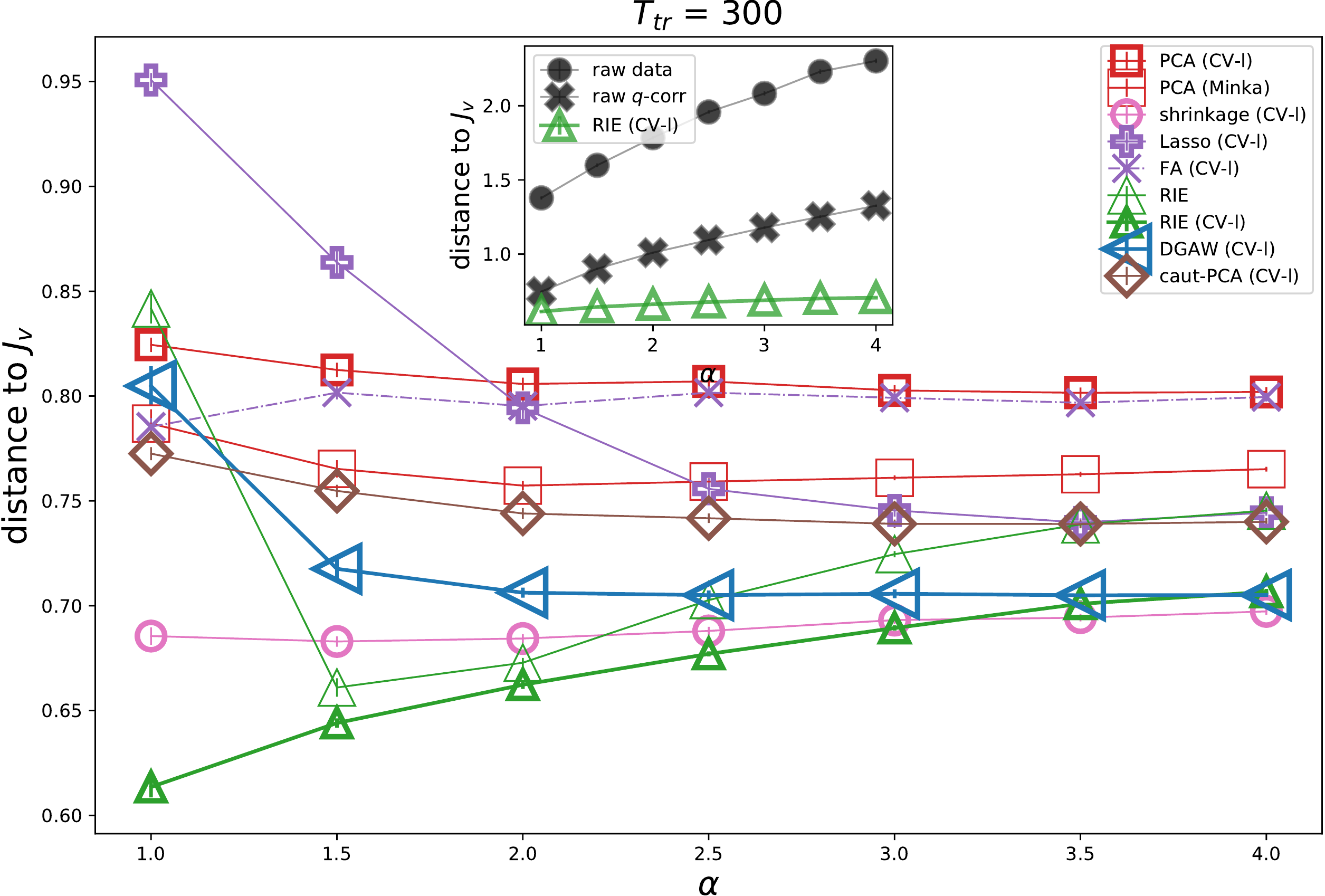} \\
\includegraphics[width=.9\columnwidth]{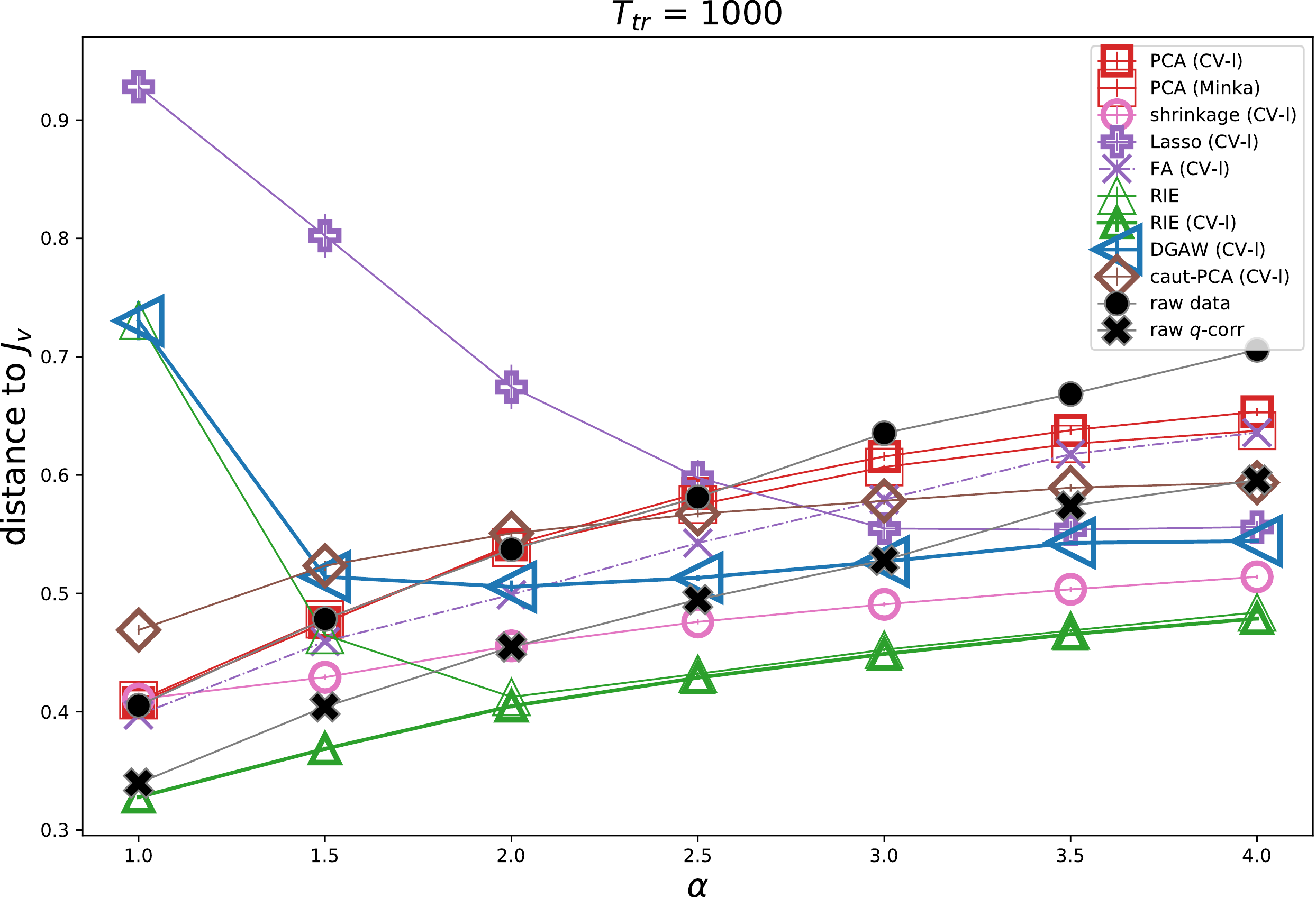} 
	\caption{ \label{fig:diffJtrue} $d(\Jv,J)$ (lower is better) versus $\alpha_\text{D}$ in synthetic data. Points and error bars are averages and standard errors of the mean across subjects, and each curve corresponds to a different method. Higher, middle and lower panel correspond to $\Ttr=144,300,1000$, respectively. In the two highest panels, the inset compare the raw, raw (q-corr.) and RIE estimators, while in the lower panel all estimators are in the main figure.}
\end{figure}

\begin{figure}[h]
\includegraphics[width=.9\columnwidth]{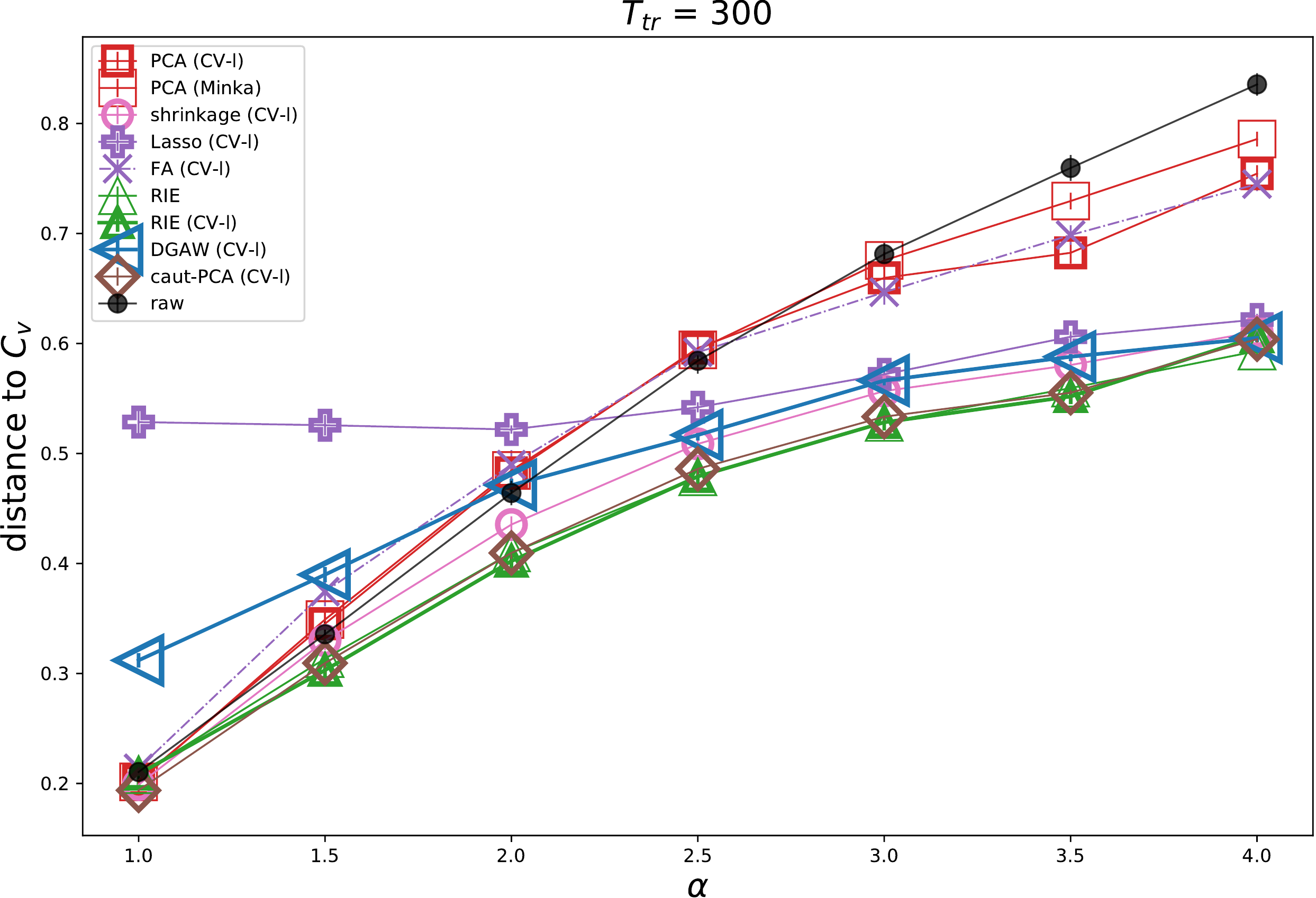} \\
	\caption{ \label{fig:diffCtrue} $d(\Cv,C)$ (lower is better) versus $\alpha_\text{D}$ in synthetic data, for $\Ttr=300$. Points and error bars are averages and standard errors of the mean across subjects, and each curve corresponds to a different method. }
\end{figure}

\begin{figure}[h]
\includegraphics[width=.9\columnwidth]{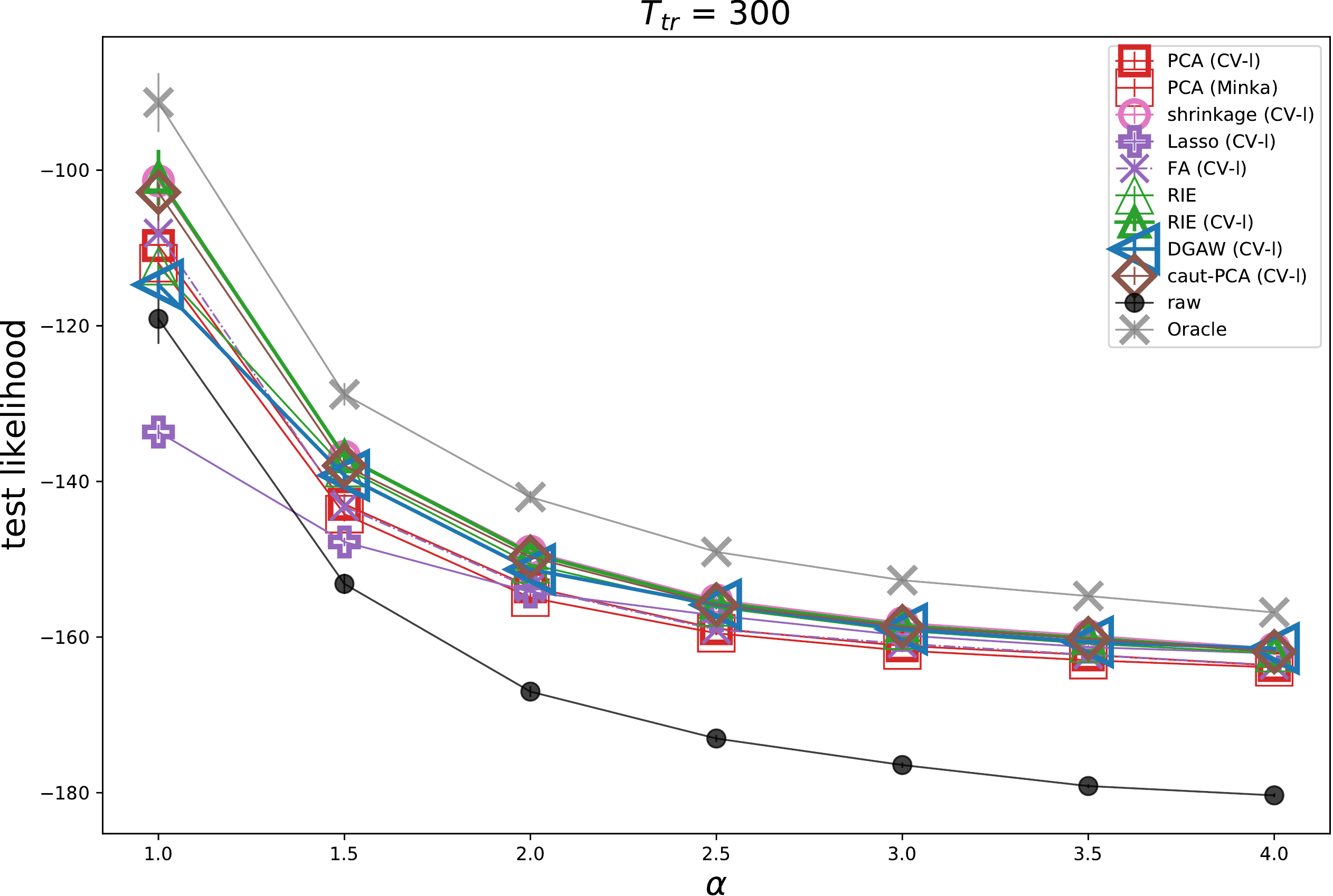} \\
	\caption{ \label{fig:likelihood} Test-likelihood $\ell$ (higher is better) versus $\alpha_\text{D}$ in synthetic data, for $\Ttr=300$. Points and error bars are averages and standard errors of the mean across subjects, and each curve corresponds to a different method.  }
\end{figure}

\begin{figure}[h]
\includegraphics[width=.9\columnwidth]{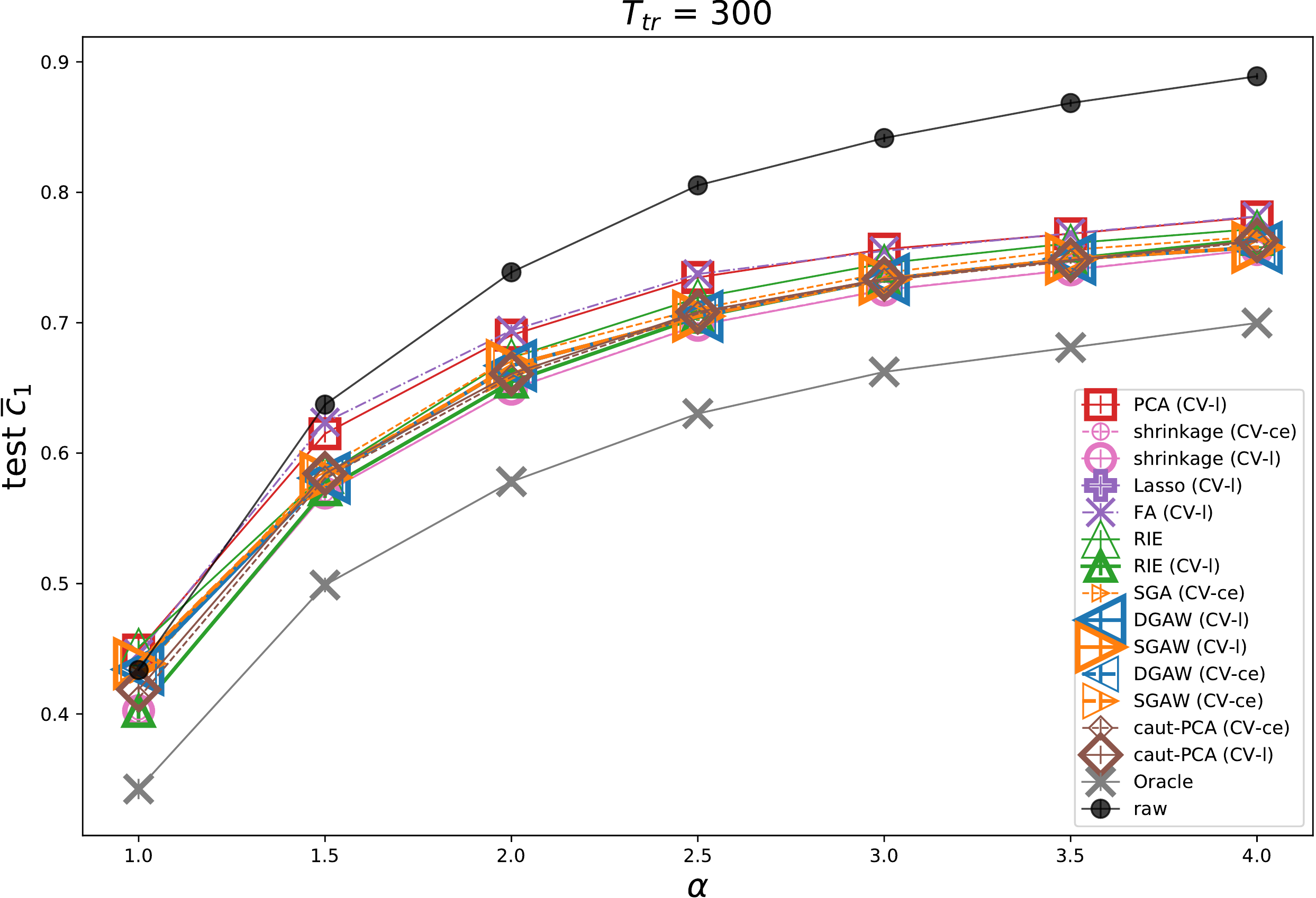} \\
	\caption{ \label{fig:completionerror} Completion error $\bar c$ (lower is better) versus $\alpha_\text{D}$ in synthetic data, for $\Ttr=300$. Points and error bars are averages and standard errors of the mean across subjects, and each curve corresponds to a different method. }
\end{figure}

\begin{figure}[h]
\includegraphics[width=.9\columnwidth]{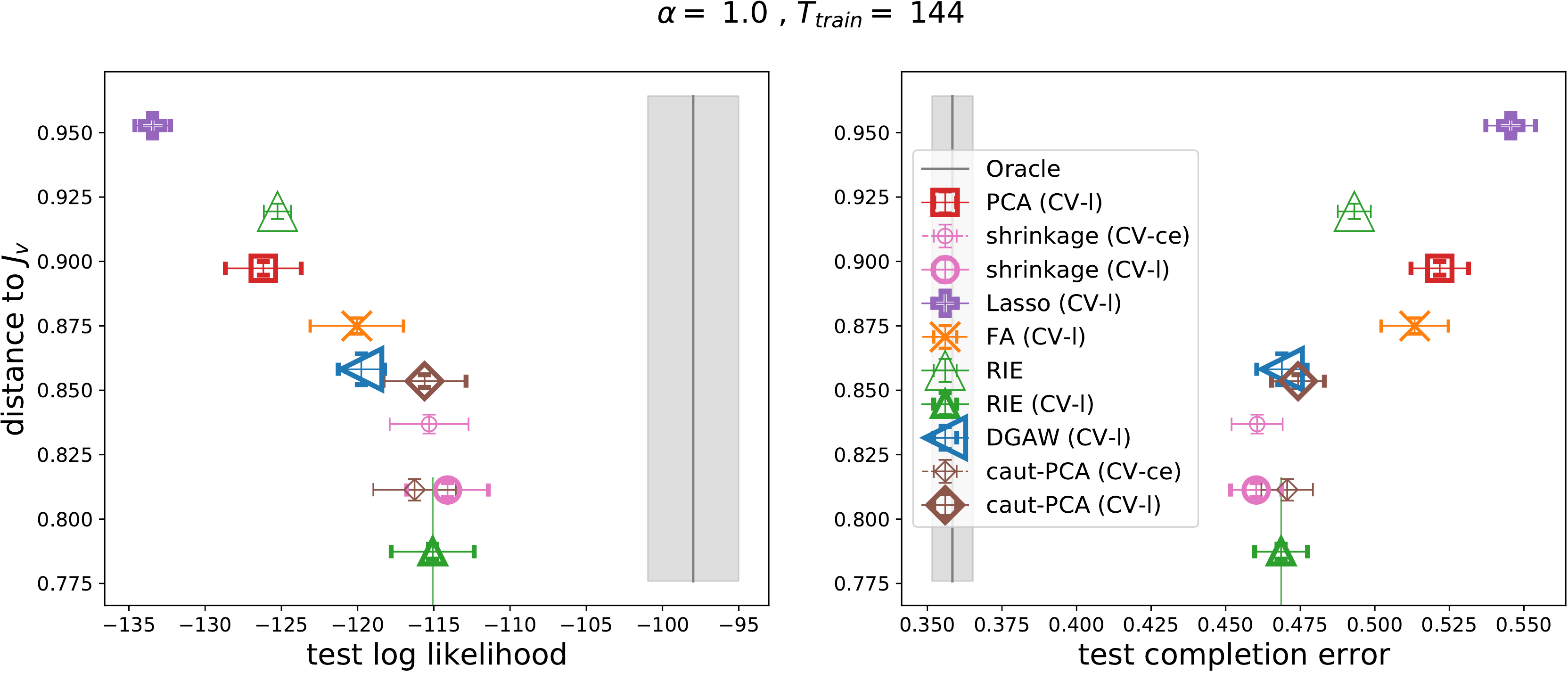} 
\includegraphics[width=.9\columnwidth]{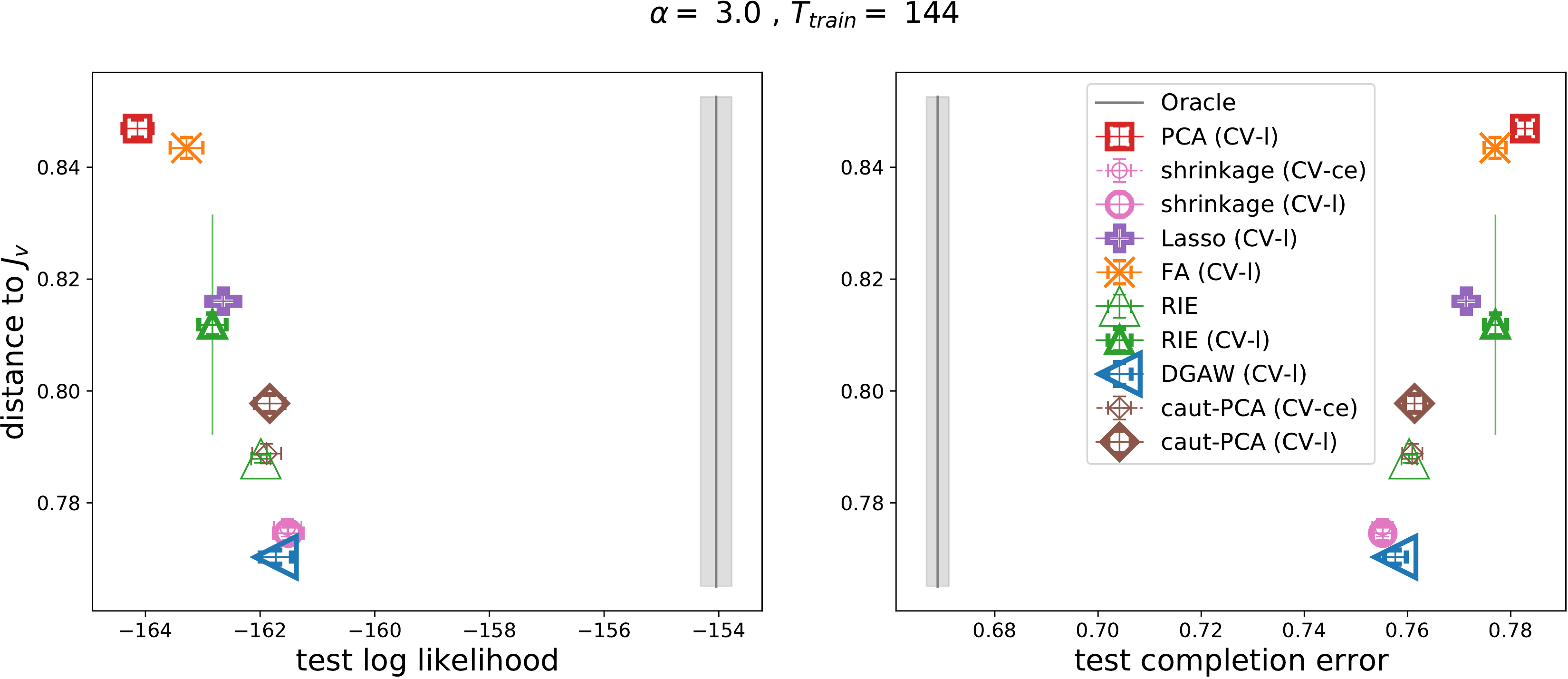} 
	\caption{ \label{fig:observable_scatter_T144} Top left: scatter plot of $d(\Jv,J)$ (lower is better) versus test-likelihood $\ell$ (higher is better), for $\Ttr=144$, $\alpha_\text{D}=1$. Top right: idem but $d(\Jv,J)$ versus $\bar c$ (lower is better). Bottom, left and right: idem, but for $\alpha_\text{D}=3$.}
\end{figure}

\begin{figure}[h]
\includegraphics[width=.9\columnwidth]{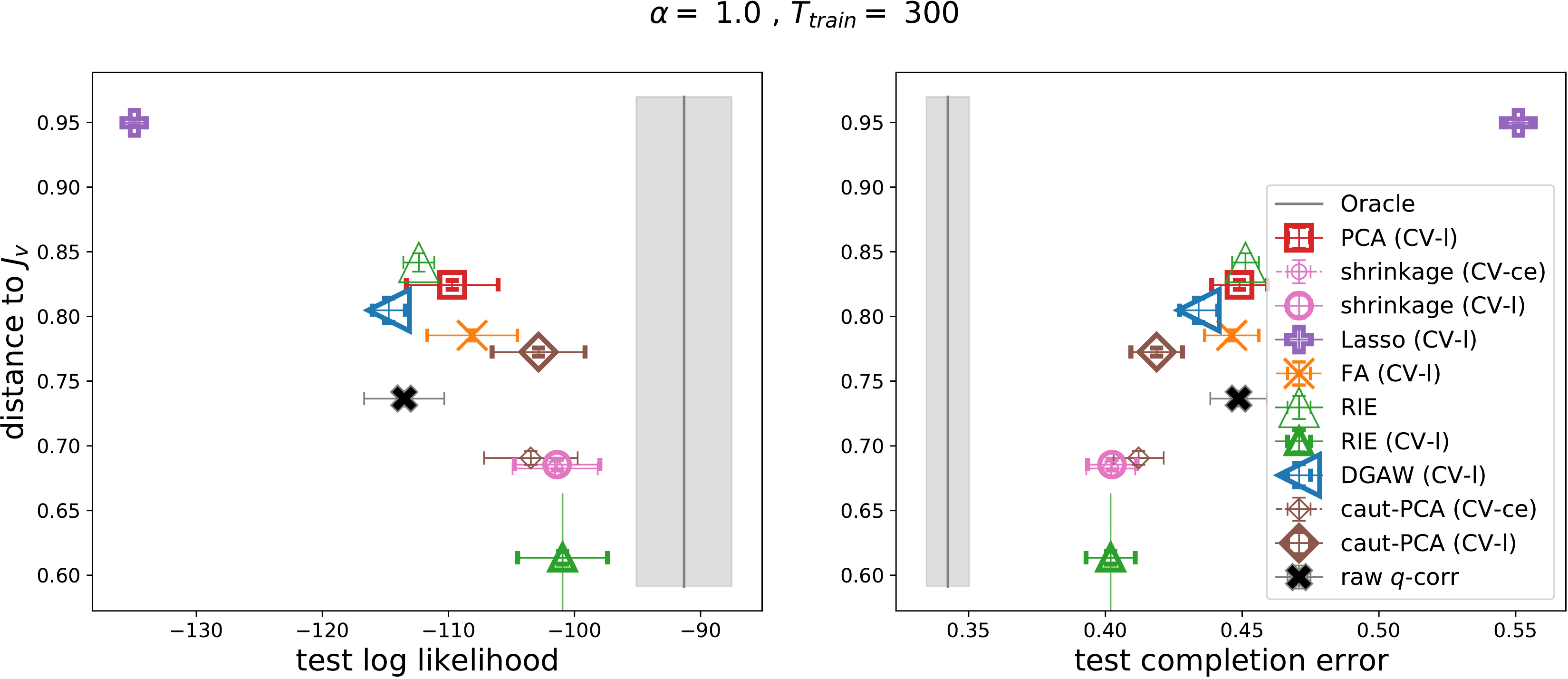} 
\includegraphics[width=.9\columnwidth]{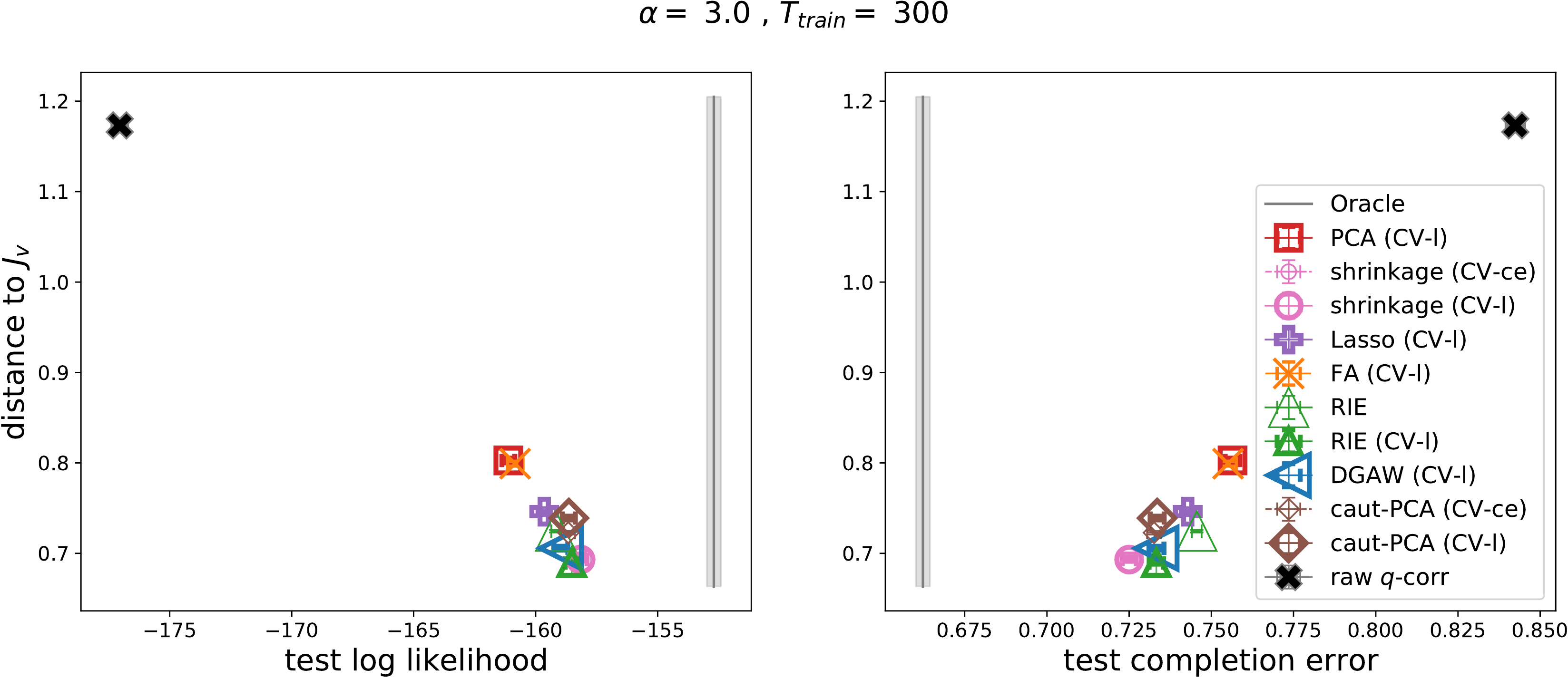} 
	\caption{ \label{fig:observable_scatter_T300} As in figure \ref{fig:observable_scatter_T144} but for $\Ttr=300$.}
\end{figure}

We here present some complementary results of the synthetic simulations. In figure \ref{fig:diffJtrue} we show $d$ versus $\alpha_\text{D}$ for various algorithms, and different values of $\Ttr$ (in different panels). This is a different perspective of the same data of figures \ref{fig:observable_scatter_dirichlet1},\ref{fig:observable_scatter_dirichlet2}, but for more values of $\alpha_\text{D}$. We show $d(\Cv,C)$ versus $\alpha_\text{D}$ for a single value of $\Ttr=300$ in figure \ref{fig:diffCtrue}. Inter-method differences are less significant, with respect to their statistical errors, than for $d(\Jv,J)$. The same occurs with the test-likelihood (figure \ref{fig:likelihood}) and the test-completion error (figure \ref{fig:completionerror}). Figures \ref{fig:observable_scatter_T144}, \ref{fig:observable_scatter_T300} present the likelihood versus the completion error for $\Ttr=300$, $\alpha_\text{D}=1,3$, respectively.

\section{Early-stopping Gradient Ascent algorithms \label{sec:GA}}

We now describe the GAW (Gradient Ascent algorithm in Wishart form) algorithm. We perform a gradient ascent search of $\ln \N(\Xin|C)$ on the $N\times N$ real matrix $Y$, defined such that $C^{-1}=Y Y^\dag$. Performing the gradient ascent search on $Y$ guarantees the positive-definiteness of the precision matrix $Y Y^\dag$ at each iteration. One first takes an initial condition in the first $\tau=0$ step, $Y(0)=1_N$, afterwards we follow the simple gradient iteration for the $rs$-element of matrix $Y$:
\begin{align}
	Y_{rs}(\tau+1) - Y_{rs}(\tau) = \eta_\text{GA}  \left. \frac{\partial}{\partial Y'_{rs}}\right|_{Y(\tau)} \ln {\N}(\Xin|C) 
	\label{eq:DGA}
\end{align}
where $C=(Y' Y'^\dag)^{-1}$ and where the {\it learning rate} $\eta_\text{GA}$ is a small, positive parameter. The gradient in equation \ref{eq:DGA} takes the form:

\begin{align}
	\frac{\partial}{\partial Y'_{rs}} \ln {\N}(\Xin|C) = \\
	=  (CY' + (CY')^\dag)-(EY' + (EY')^\dag )_{rs} 
	\label{eq:gradient}
\end{align}
where $E$ is the unbiased estimator of the covariance matrix given $\Xin$. The term $C Y$ may be computed from $Y$ using the singular value decomposition $Y=W \Lambda^{-1/2} Z$, where $W$, $Z$ are unitary matrices and $\Lambda$ is the diagonal eigenvalue matrix of $C$, afterwards taking $C Y=W\Lambda^{1/2} Z$. The iterations stop when the quality criterion $Q(\Xva|C(\tau))$ decreases from the $\tau$-th to the $\tau+1$-th iteration, and the cleaned correlation matrix is taken as $C(\tau)$.

The Stochastic Gradient Ascent algorithm is based on the above described (deterministic) GA algorithm but, instead of the deterministic gradient ascent of equation \ref{eq:DGA}, we use a stochastic gradient ascent rule, inspired in artificial neural network learning:

\begin{align}
	&Y(\tau+1) - Y(\tau) = \eta \left( M(\tau) + M^\dag(\tau) \right) \\
	&M(\tau) = ( C(\tau)  -E^{(\tau)} )  Y(\tau)  
	\label{eq:SGA}
\end{align}
where $\E^{(\tau)}$ is a random bootstrapping of the sample correlation matrix, consisting in the covariance matrix of a subset of $B\le T$ sample vectors composing the training set, with repeating indices. In other words, at each iteration of the gradient ascent algorithm, the sample term of the gradient in eq. \ref{eq:DGA} is not constant, but calculated with a random bootstrapping of the data, different at each epoch. In this case, the stopping criterion is consequently modified: the iterations stop when $Q(\Xva|Q(\tau))$ decreases for $\tau_\text{d}$ consecutive iterations.

{\bf Adding the constant trace constraint.} Suppose that $Y^*$ is the solution satisfying $Y^*=\arg\max[\N(\Xin|(Y Y^\dag)^{-1})]$ subject to the constraint $\traza(C)=N$ with $C=(Y Y^\dag)^{-1}$. Then, the $Y^*$ satisfies:

\begin{align}
	&\left.	\partial_Y\left[ \ln \N(\Xin|C) -\mu\,[\traza(C)-N] \right] \right|_{Y^*,\mu^*} =0 \\
	&\left.	\partial_\mu\left[ \ln \N(\Xin|C) -\mu\,[\traza(C)-N] \right] \right|_{Y^*,\mu^*} =0 
\end{align}
where $\mu$ is the Lagrange multiplier associated to the constraint. We have, hence, a further scalar variable, and a further equation in the satisfaction problem (that we solve only approximately, since we apply the early stopping criterion). The two above coupled equations induce the following Euler iterative dynamics in the variables $Y$, $\mu$:

\begin{align}
	\mu(\tau+1)-\mu(\tau)&=\pm\eta_\mu \, \left[ N-\traza(C(\tau)) \right] \\ 
	Y(\tau+1)-Y(\tau)&=\eta_\text{GA}\,\left[ M(\tau) + M^\dag(\tau) \right]\\
	M(\tau)&:=C(\tau)Y(\tau)-E Y(\tau) +2 \mu(\tau)\, C^{2}(\tau) Y(\tau) 
\end{align}
where $\eta_\mu$ is the learning rate associated to the updating of $\mu$, that we set constant and larger than $\eta_\text{GA}$ (in the numerical calculations, we actually set $\eta_\mu=10\eta_\text{GA}$). 

\begin{figure}[h]
\includegraphics[width=.9\columnwidth]{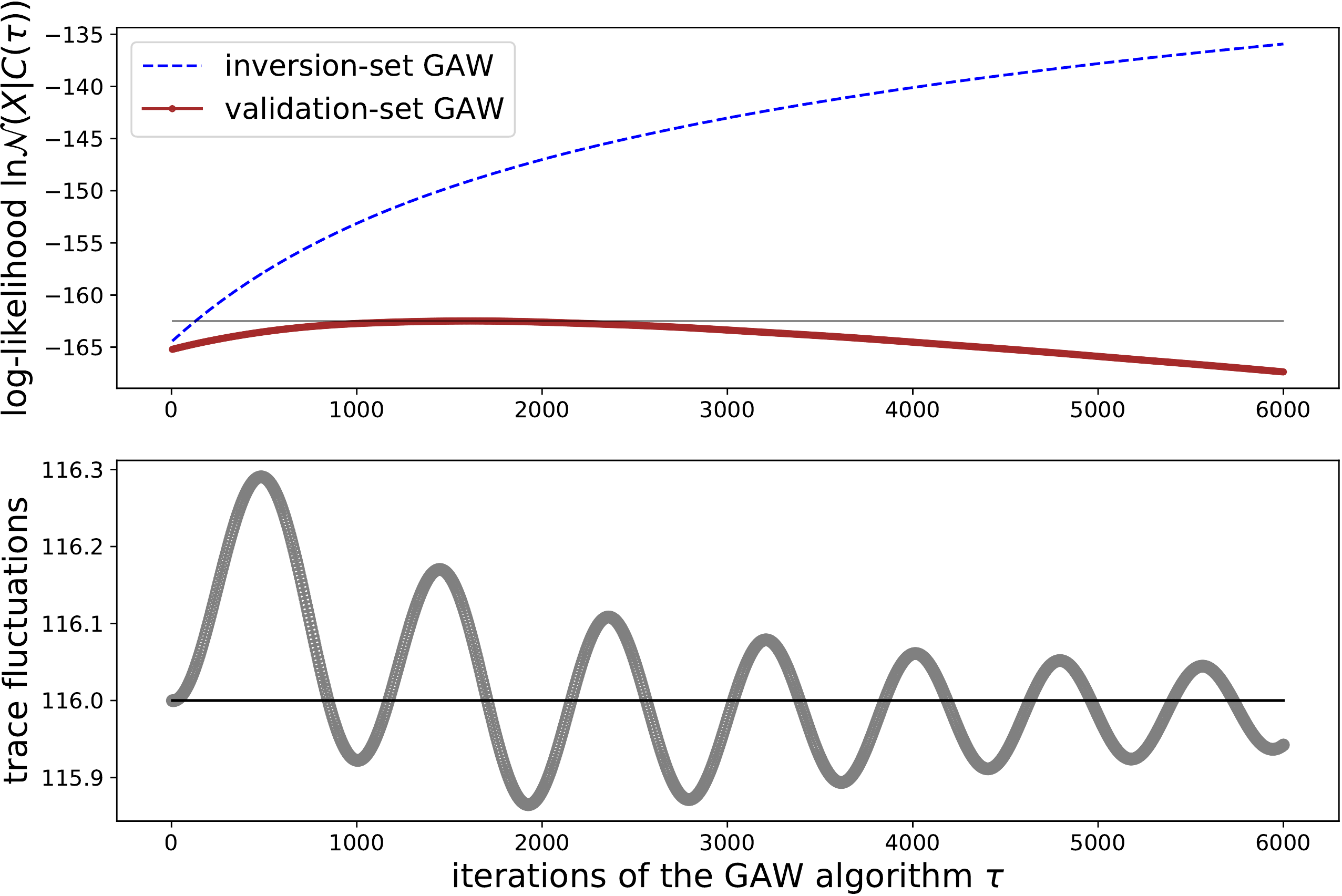} 
	\caption{ \label{fig:trace_oscillations} Illustration of the GAW algorithm. Upper panel: validation- and inversion-set likelihoods versus the number of iterations $\tau$. Lower panel: the trace of the resulting covariance matrix $\traza(C(\tau))$, versus $\tau$. In this example, the database $X$ is a synthetic dataset with $\alpha_\text{D}=3$, $\Ttr=144$, $\Tin=(5/6)\Ttr$. The algorithm parameters are: $\eta_\text{GA}=10^{-4}$, $\eta_{\lambda}=5\,10^{-2}$.}
\end{figure}

We show the validation- and inversion-likelihood as a function of the number of iterations in the GAW algorithm, in figure \ref{fig:trace_oscillations}. When the validation-set likelihood reaches its maximum value (horizontal line in figure \ref{fig:trace_oscillations}), the iterations stop and the resulting $C$ is taken as the regularised matrix. The lower panel of the figure shows the behaviour of $\traza(C(\tau))$ and its oscillations around its required value $N=116$. Increasing the value of $\eta_\lambda$ reduces the amplitude of the oscillations in $\traza(C)$, but this does not have a statistically significant impact on the results for the subject-averaged values of $d$, $\ell$.  

\section{Details of the numerical simulations \label{sec:details}}

We present some details of the numerical algorithms (see \cite{repository}). The array of values of the cross-validated hyper-parameters is $p=1,\ldots,N-1$ for PCA and caut-PCA; for FA, the hyper-parameter $r$ takes the same values; for shrinkage,  $\alpha \in \{1-10^{x}\}$ with $x$ taking $30$ equally spaced values between $-2$ and $-0.1$; for RIE, $\eta \in {x\, N^{-1/2}}$ with $x$ taking the values $0.1, 0.2, 0.5, 1, 2, 5, 10, 20, 50, 100$. For the Lasso algorithm, we have employed the {\sf scikit-learn} implementation \cite{scikit-learn}, called {\sf GraphicalLassoCV}, with an initial $4$-length grid with $4$ refinements and $1000$ maximum number of iterations. Also for FA and shrinkage we use the {\sf scikit-learn} versions. For the GA and GAW algorithms we employ: $\eta_\text{GA}=10^{-4}$ and $\eta_\mu=10^{-2}$. For the stochastic version, SGAW, we employ a batch size $B=\Tin/4$.

\end{document}